\documentclass[11pt]{article}
\usepackage[a4paper]{geometry}
\usepackage[standard]{ntheorem}
\newcommand{\keywords}[1]{\def\and{, }
{\bf Keywords}: #1}
\newcommand{\email}[1]{\texttt{#1}}

\usepackage{microtype}
\usepackage{graphicx}
\usepackage{amsmath}
\usepackage{amssymb}
\usepackage[ruled]{algorithm}
\usepackage{multirow}
\usepackage{float}

\usepackage{mathabx}
\usepackage{xspace}
\usepackage{hyperref}

\usepackage{tikz}
\usetikzlibrary{arrows}

\def\qed {{                
   \parfillskip=0pt        
   \widowpenalty=10000     
   \displaywidowpenalty=10000  
   \finalhyphendemerits=0  
   \leavevmode             
   \unskip                 
   \nobreak                
   \hfil                   
   \penalty50              
   \hskip.2em              
   \null                   
   \hfill                  
   $\square$
   \par}}                  

\newcommand{\sep}{\mathrel{|}}

\newcommand{\true}{\ensuremath{\mathbf{true}}}
\newcommand{\false}{\ensuremath{\mathbf{false}}}
\newcommand{\Mgr}{\ensuremath{\mathit{Mgr}}}
\newcommand{\Emp}{\ensuremath{\mathit{Emp}}}
\newcommand{\John}{\ensuremath{\mathit{John}}}
\newcommand{\Mary}{\ensuremath{\mathit{Mary}}}

\newcommand{\Ken}{\ensuremath{\mathit{Ken}}}
\newcommand{\Rep}{\mathit{Rep}}
\newcommand{\XRep}{\mathcal{X}\!\mathit{Rep}}
\newcommand{\YRep}{\mathcal{Y}\!\mathit{Rep}}
\newcommand{\CRep}{\mathcal{C}\!\mathit{Rep}}
\newcommand{\PRep}{\mathcal{P}\!\mathit{Rep}}
\newcommand{\GRep}{\mathcal{G}\!\mathit{Rep}}

\newcommand{\step}[1]{{\tt \footnotesize #1}:}

\begin{document}
\title{Prioritized Repairing and Consistent Query Answering in
  Relational Databases\thanks{Research partially supported by NSF
    grants IIS-0119186 and IIS-0307434, Ministry of Higher Education
    and Research, Nord-Pas de Calais Regional Council and FEDER
    through the' Contrat de Projets Etat Region (CPER) 2007-2013,
    Enumeration project ANR-07-blanc, and Polish Ministry of Science
    and Higher Education research project N N206 371339.}}

\author{
  S\l{}awek Staworko\thanks{Corresponding author. Part of this
    research was done when the author was a PhD student at the
    University at Buffalo.}\\
  Mostrare project\\
  INRIA \& LIFL (CNRS UMR8022)\\
  \email{slawomir.staworko@inria.fr} 
  \and Jan Chomicki\\
  Computer Science and Engineering\\
  University at Buffalo\\
  \email{chomicki@buffalo.edu} 
  \and Jerzy Marcinkowski\\
  Institute of Informatics\\
  Wroc\l{}aw University \\
  \email{jma@cs.uni.wroc.pl} 
}

\maketitle
\thispagestyle{empty}
\begin{abstract}
  A consistent query answer in an inconsistent database is an answer
  obtained in every (minimal) repair. The repairs are obtained by
  resolving all conflicts in all possible ways. Often, however, the
  user is able to provide a preference on how conflicts should be
  resolved. We investigate here the framework of \emph{preferred
    consistent query answers}, in which user preferences are used to
  narrow down the set of repairs to a set of \emph{preferred
    repairs}. We axiomatize desirable properties of preferred
  repairs. We present three different families of preferred repairs
  and study their mutual relationships. Finally, we investigate the
  complexity of preferred repairing and computing preferred consistent
  query answers. 

\end{abstract}
\section{Introduction}
In many novel database applications, violations of integrity
constraints cannot be avoided. A typical example is integration of two
consistent data sources that contribute conflicting information.
Inconsistencies also often occur in the context of long-running
operations where transaction mechanisms are not employed. Finally,
integrity enforcement may be disabled because of efficiency
considerations.  Integrity constraints, however, capture important
semantic properties of the stored data. These properties directly
influence the way a user formulates a query. Evaluation of the query
over an inconsistent database may yield answers that are meaningless
or misleading.

The framework of \emph{repairs} and \emph{consistent query answers}
\cite{ArBeCh99} has been proposed to offset the impact of
inconsistencies on the accuracy of query answers. A \emph{repair} is a
consistent database minimally different from the given one, and a {\em
  consistent answer} to a query is an answer present in \emph{every}
repair. This approach does not physically remove any facts from the
database. The framework of \cite{ArBeCh99} has served as a foundation
for most of the subsequent work in the area of querying inconsistent
databases (for the surveys of the area see
\cite{BeCh03,Be06,ChMa05,Ch07,Be11}, other works include
\cite{Wij05,Wij10}).

Recently, the problem of database repairing has received an enlivened
interest~\cite{AfKo09,Fa08}. Essentially, the goal is to construct a
repair of a possibly inconsistent instance by resolving every conflict
present in the given instance. In the case of denial constraints, the
class of constraints we consider in this paper, a conflict is simply a
set of facts that are present in the given instance that together
violate a constraint. A resolution of a conflict is the deletion of
one of the facts creating the conflict. Typically, there exists more
than one repair and a repairing algorithm needs to make some
nondeterministic choices when repairing the database instance. It is
desirable for the algorithm to be \emph{sound} i.e., always producing a
repair, that is, an instance which is not only consistent but also
minimally different from the given one. It is even more desirable for
the algorithm to be \emph{complete} i.e.,  allowing to produce every
repair, with an appropriate sequence of choices~\cite{StCh08}.
\begin{example}\label{ex:repairs}
Consider the schema consisting of two relations 
\[
\Emp(Name, Salary, Dept)
\quad\text{and}\quad
\Mgr(Name, Salary, Dept),
\] 
and the set of constraints $F_0$ consisting of
\begin{align*}
&\Emp : Name \rightarrow Name\,\, Salary\,\, Dept,\\
&\forall x,y,z,x',y'.\  \neg [ \Emp(x,y,z) \land \Mgr(x',y',z) \land y > y'].
\end{align*}
The first constraint is a key dependency requiring the employee
information to be associated with her name. The second constraint is a
denial constraint requiring that no employee of a department earns
more than the manager of the department.

Now, consider the inconsistent database instance 
\begin{align*}
I_0=\{
& \Emp(\John, \$40k, IT), \Emp(\John, \$50k, IT), \\
& \Emp(\John, \$80k, IT), \Mgr(\Mary, \$70k, IT)
\}.
\end{align*}
This instance contains three conflicts w.r.t.\ the functional
dependency and one conflict w.r.t.\ the denial constraint. $I_0$ has
three repairs w.r.t.\ $F_0$:
\begin{align*}
& I_1' = \{ \Emp(\John,\$80k,IT)\},\\
& I_2' = \{ \Emp(\John,\$50k,IT), \Mgr(\Mary,\$70k,IT)\},\\
& I_3' = \{ \Emp(\John,\$40k,IT), \Mgr(\Mary,\$70k,IT)\}.
\end{align*}
Consider the query $Q_0=\exists x, y .\  \Emp(\John, x, y) \land x >
\$60k$ asking whether $\John$ earns more than $\$60k$. The answer to
$Q_1$ in the database instance $I_0$ is $\true$. However, $\true$ is
not a consistent answer to $Q_1$ because of the repairs $I_2'$ and
$I_3'$. \qed
\end{example}

One of the drawbacks of the framework of consistent query answers is
that it considers all possible ways to resolve the existing
conflicts. The user, however, may have a preference on what
resolutions to consider. Typical information used to express the
preference includes:
\begin{itemize}
\itemsep 0pt
\item the timestamp of creation/last modification of the fact; the
  conflicts can be resolved by removing from consideration old,
  outdated facts,
\item the source of the fact (in data integration setting); the user
  can consider the data from one source more reliable than the data
  from another source,
\item the data values stored in the conflicting facts.
\end{itemize}
To improve the quality of consistent answers we propose extending the
framework of repairs and consistent query answers with the preference
information. We use the preference information to define a set of {\em
  preferred} repairs (a subset of all repairs). Query answers obtained
in every preferred repair are called \emph{preferred consistent query
  answers}. For instance, in the previous example if the database
contains an employee who earns more than her manager, then we might
prefer to remove the information about the employee rather than the
information about the manager of the department. Then the preferred
repairs are $I_2'$ and $I_3'$, and consequently, $\false$ is the
preferred consistent answer to $Q_0$.

We observe, however, that there may be more that one way to select the
preferred repairs based on the user preference; especially, when a
resolution of one conflict affects the way in which another conflict
can be resolved.
\begin{example}\label{ex:preference}
We take the schema consisting of one relation name 
\[
\Mgr(Name, Salary, Dept)
\] 
with two functional dependencies 
\[
\Mgr : Name\rightarrow Salary\,\, Dept
\quad\text{and}\quad
\Mgr : Dept \rightarrow Name\,\, Salary. 
\]
Consider the following inconsistent instance
\begin{align*}
I_1 = \{
& \Mgr(Bob,\$70k,RD), \Mgr(\Mary,\$40k,IT), \Mgr(\Ken,\$60k,IT), \\
& \Mgr(Bob,\$60k,AD), \Mgr(\Mary,\$50k,PR), \Mgr(\Ken,\$50k,PR)
\}
\end{align*}
This instance contains five conflicts:
\begin{enumerate}
\item $\Mgr(Bob,\$70k,RD)$ and $\Mgr(Bob,\$60k,AD)$.
\item $\Mgr(\Mary,\$40k,IT)$ and $\Mgr(\Mary,\$50k,PR)$,
\item $\Mgr(\Ken,\$60k,IT)$  and $\Mgr(\Ken,\$50k,PR)$,
\item $\Mgr(\Mary,\$40k,IT)$ and $\Mgr(\Ken,\$60k,IT)$,
\item $\Mgr(\Mary,\$50k,PR)$ and $\Mgr(\Ken,\$50k,PR)$,
\end{enumerate}
These conflicts may arise from changes that are not yet fully
propagated. For instance, $Bob$ may have been moved to manage $R\&\!D$
department while previously being the manager of $AD$, or $Bob$ may
have been moved from $AD$ department to $RD$ department. Similarly,
$\Mary$ may have been promoted to manage $PR$ whose previous manager
was moved to manage $IT$, or conversely, $\John$ may have been moved to
manage $IT$, while $\Mary$ was moved from $IT$ to $PR$.

The set of repairs of $I_1$ consists of four instances:
\begin{align*}
& I_1'=\{\Mgr(Bob,\$70k,RD),\Mgr(\Mary,\$50k,PR),\Mgr(\Ken,\$60k,IT)\},\\
& I_2'=\{\Mgr(Bob,\$70k,RD),\Mgr(\Mary,\$40k,IT),\Mgr(\Ken,\$50k,PR)\},\\
& I_3'=\{\Mgr(Bob,\$60k,AD),\Mgr(\Mary,\$40k,IT),\Mgr(\Ken,\$50k,PR)\},\\
& I_4'=\{\Mgr(Bob,\$60k,AD),\Mgr(\Mary,\$50k,PR),\Mgr(\Ken,\$60k,IT)\}.
\end{align*}
Suppose that the user prefers to resolve a conflict created by two
facts referring to the same person by \emph{removing the tuples with
  the smaller salary}. This preference expresses the belief that if a
manager is being reassigned, her salary is not decreased.  It applies
to the first conflict: the fact $\Mgr(Bob,\$70k,RD)$ is preferred over
$\Mgr(Bob,\$60k,AD)$. Similarly, the preference applies to the second
and the third conflict. It does not apply to the last two conflicts as
each of them involves facts referring to different persons.

The preference information on resolutions of the first conflict allows
us to eliminate the last two repairs $I_3'$ and $I_4'$. Similarly, by
applying the preference to the conflicts 2 and 3 we may also eliminate
the repair $I_2'$. This leaves us with only one preferred repair
$I_1'$.

We observe that while the preference applies to conflicts 1, 2, and 3,
it does not apply to conflicts 4 and 5 because conflicts 4 and 5
involve facts about different persons. However, the preferential
resolution of conflicts 2 and 3 \emph{implicitly} resolves the
conflicts 4 and 5, which may not be desirable. Consequently, one may
find the reasons for eliminating $I_2'$ insufficient. \qed
\end{example}
In this paper we consider three different families of \emph{preferred
  repairs}. The families are based on various notions of compliance of
a repair with the user preference. The first two notions, global and
Pareto optimality, check if the compliance of a repair $I'$ can be
improved by replacing a subset of facts $X\subseteq I'$ with a {\em
  more preferred} subset of facts $Y\subseteq I\setminus I'$. These
notions differ in the way they lift preference on facts to preferences
on sets of facts.

\emph{Global optimality} requires that for every element in $X$ there
is a more preferred element in $Y$. This approach is inspired by the
work on preferential reasoning~\cite{Ha97} and corresponds to the
first way of selecting preferred repairs in the previous example. For
instance, $I_2'$ is not globally optimal because we can replace
$X=\{\Mgr(\Mary,\$40k,IT), \Mgr(\Ken,\$50k,PR)\}$ with a more
preferred $Y=\{\Mgr(\Mary,\$50k,PR), \Mgr(\Ken,\$60k,IT)\}$, obtaining
the only globally-optimal instance $I_1'$.

\emph{Pareto optimality} requires a stronger support from the
preference to conclude that the compliance of a repair with the
preference can be improved: every element of $Y$ needs to be preferred
over every element of $X$. This approach is inspired by the
construction of the Pareto-optimal set of vectors~\cite{KoPa07} and it
corresponds to the second way of selecting preferred repairs in the
previous examples. For instance, $I_3'$ is not Pareto optimal because
we can replace $X=\{\Mgr(Bob,\$60k,AD)\}$ with
$Y=\{\Mgr(Bob,\$70,RD)\}$. We remark that for this notion of
optimality the compliance of $I_2'$ with the preference cannot be
further improved, thus $I_2'$ is Pareto optimal.

The third notion of \emph{completion optimality} uses a different approach
to verify an optimal compliance of a repair with the preference. It
views the preference only as a step towards a \emph{total} preference
i.e., preference that specifies the preferred resolution of every
conflict, which yields exactly one repair. A repair is completion optimal
if the preference can be extended to a total preference that yields
the given repair. In the previous example completion optimality coincides
with global optimality. The instance $I_1'$ is completion optimal because
we can add an appropriate preference for conflicts 4 and 5. 

For every family of preferred repairs we present a repairing
algorithm. Each of them is \emph{sound} i.e., it produces a repair
belonging to the corresponding family of preferred repairs, and {\em
  complete} i.e., every repair from the family of preferred repairs
can be constructed using the corresponding repairing algorithm. For
the family of globally-optimal repairs and the family of
Pareto-optimal repairs we define two pre-orders on repairs whose
maximal elements are exactly the globally-optimal repairs and
Pareto-optimal repairs respectively. It is an open question whether
such an order can be defined for completion-optimal repairs.

We also adapt two basic decision problems: \emph{repair
  checking}~\cite{ChMa05,AfKo09} and \emph{consistent query
  answering}~\cite{ArBeCh99} to obtain preferred repair checking and
preferred consistent query answering. Basically, \emph{preferred repair
  checking} is finding if a given database instance is a preferred
repair, and \emph{preferred consistent query answering} is finding if
an answer to a query is obtained in every preferred repair.

Recall from~\cite{ChMa05} that the class of denial constraints
lies on the tractability frontier of consistent query answering. 
On the one hand for the class of denial constraints repair
checking and computing consistent answers to quantifier-free ground
queries is in PTIME. On the other hand, computing consistent answers
to conjunctive queries i.e., conjunctions of positive literals with
existential quantifiers, becomes coNP-complete even in the presence of
one functional dependency i.e., a simple denial constraint. It seems
natural that this tractability frontier should shift after
incorporating a nontrivial component into the inputs of the definitions 
of the decision problems and the interesting question is how much.

We show that using the notion of global optimality leads to
intractability of both preferred consistent query answering, which
becomes $\Pi_2^p$-complete, and preferred repair checking, which
becomes coNP-complete. The complexity is reduced if we use the notion
of Pareto optimality: the preferred consistent query answering becomes
coNP-complete and preferred repair checking is in LOGSPACE. Using
completion-optimal repairs also reduces the complexity: preferred repair
checking is in PTIME and preferred consistent query answering becomes
coNP-complete. It is an open question whether in this case the
preferred repair checking is PTIME-complete or in LOGSPACE. Finally,
we identify a tractable case of quantifier-free ground queries and one
FD per relation, for which preferred consistent query answering is in
PTIME for every of the aforementioned families of preferred repairs.

The contributions of this paper are:
\begin{itemize}
\item A formal framework of families of preferred repairs and
  preferred consistent query answers for relational databases.
\item A list of desirable properties of families of preferred repairs.
\item Three different families of preferred repairs based on different
  notions of optimal compliance with the user preference.
\item Repairing algorithm for every family of preferred repairs. The
  algorithms are both sound and complete.
\item A thorough analysis of computational implications of preferences
  in the context of repairing and consistent query answers.
\end{itemize}
The presented work is an extension of~\cite{StChMa06}. The current
paper extends the framework of preferred consistent query answers to
denial constraints (instead of functional dependencies), provides
detailed proofs of all claims, and presents sound and complete
repairing algorithms for every considered family of preferred repairs
(instead of just the repairing algorithm for the family of
completion-optimal repairs only). Additionally, we further broaden the
analysis of computational complexity by identifying a family of
preferred repairs for which preferred repair checking is in LOGSPACE,
offering a possibility of parallel implementation for this decision
problem.

The paper is organized as follows. In Section~\ref{sec:basic} we
recall basic notions of relational databases and the framework of
repairs and consistent query answers. In Section~\ref{sec:preferences}
we extend this framework with preferences on conflict resolution. In
Sections~\ref{sec:global},~\ref{sec:pareto}, and~\ref{sec:common} we
present the families of globally-, Pareto-, and completion-optimal repairs
respectively. We investigate their properties and mutual
relationships, and analyze the computational implications of their
semantics. In Section~\ref{sec:complexity} we present a tractable case
of preferred consistent query answering. Section~\ref{sec:related}
contains a discussion of related work. Finally, in
Section~\ref{sec:future} we summarize our results and outline
directions for future work.
\section{Preliminaries}
In this section we recall the basic notions of relational
databases~\cite{AbHuVi95} and the framework of consistent query
answers~\cite{ArBeCh99}.
\label{sec:basic}
A database \emph{schema} $\mathcal{S}$ is a set of relation names of
fixed arity (greater than $0$) whose attributes are drawn from an
infinite set of names $U$. Every element of $U$ is typed but for
simplicity we consider only two disjoint infinite domains:
$\mathsf{Q}$ (rationals) and $D$ (uninterpreted constants). We assume
that two constants are equal if and only if they have the same name,
and we allow the standard \emph{built-in} relation symbols $=$ and
$\neq$ over $D$. We also allow the built-in relation symbols $=$,
$\neq$, $<$, $\leq$, $>$, and $\geq$ with their natural interpretation
over $\mathsf{Q}$. We use these symbols together with the vocabulary
$\mathcal{S}$ of relational names to build a first-order language
$\mathcal{L}$. An $\mathcal{L}$-formula is:
\begin{itemize}
\item \emph{closed} (or a \emph{sentence}) if it has no free variables,
\item \emph{ground} if it has no variables whatsoever,
\item \emph{quantifier-free} if it has no quantifiers,
\item \emph{atomic} if it has no quantifiers and no Boolean connectives.
\end{itemize}
Finally, a \emph{fact} is an atomic ground $\mathcal{L}$-formula.

Database \emph{instances} are finite, first-order structures over the
schema. Often, we find it more convenient to view an instance $I$
as the finite set of all facts satisfied by the instance i.e.,
$\{R(t)\sep R\in \mathcal{S}, I\models R(t)\}$. 
In this paper we use the standard notion of \emph{satisfaction} (or
\emph{entailment}) of an $\mathcal{L}$-formula $\phi$ in a database
instance $I$, in symbols $I\models\phi$. An $\mathcal{L}$-formula is
\emph{valid} iff it is satisfied in every database instance $I$. Notice
that the validity of a quantifier-free ground formula using 
only built-in predicates is decided in a straightforward fashion.

In the sequel, we denote tuples of variables by
$\bar{x},\bar{y},\ldots$, tuples of constants by $t,s,\ldots$,
quantifier-free formulas using only built-in predicates by $\varphi$,
instances by $I,J,\ldots$, relation names by $R,P,\ldots$, and
attribute names by $A,B,C,\ldots$. The symbols $X,Y,\ldots$ are used
to denote finite sets of attribute names. We also use $X,Y,\ldots$ to
denote finite sets of facts, and it will always be clear from the
context which usage is employed.

\subsection{Integrity constraints}
In general, an integrity constraint is a closed
$\mathcal{L}$-formula. In this paper we consider the class of {\em
  denial constraints}, $\mathcal{L}$-sentences of the form
\begin{equation*}\label{eq:den}
\forall \bar{x}.\  \neg [
R_1(\bar{x}_1)\land \ldots \land R_n(\bar{x}_n) \land
\varphi(\bar{x})
],
\end{equation*}
where $\varphi(\bar{x})$ is a quantifier-free formula referring to
built-in relation names only and
$\bar{x}_1\cup\ldots\cup\bar{x}_n=\bar{x}$. We also make a natural
assumption that $n>0$.

The class of denial constraints contains \emph{functional
  dependencies\/} (FDs) commonly formulated as $R : X \rightarrow Y$,
where $X$ and $Y$ are sets of attributes of $R$. An FD $R:X\rightarrow
Y$ is expressed by the following denial constraint
\[
\forall \bar{x},\bar{y}_1,\bar{y}_2,\bar{z},\bar{z}' .\  \neg [ R(\bar{x},\bar{y}_1,\bar{z}) \land R(\bar{x},\bar{y}_2,\bar{z}')\land \neg (\bar{y}_1 = \bar{y}_2)],
\]
where $\bar{x}$ is the vector of variables corresponding to the
attributes $X$, and $\bar{y}_1$ and $\bar{y}_2$ are two vectors of
variables corresponding to the attributes $Y$. A \emph{key dependency}
is a functional dependency $R: X\rightarrow Y$, where $Y$ comprises
all attributes of $R$. If the relation name is known from context, for
clarity we omit it in our notation i.e., we write $X\rightarrow Y$
instead of $R:X\rightarrow Y$. Database consistency is defined in the
standard way.
\begin{definition}
  Given a database instance $I$ and a set of integrity constraints
  $F$, $I$ is \emph{consistent} with $F$ if $I\models F$ in the
  standard model-theoretic sense; otherwise $I$ is \emph{inconsistent}.
\end{definition}
We observe that an empty instance satisfies any set of denial
constraints. This conforms to the behavior of typical SQL database
management systems: an empty database satisfies any set of constraints
expressed in SQL. Also, note that denial constraints can be
represented using standard SQL assertions. We remark, however, that
the converse is not necessarily the case.

\subsection{Queries}
In this paper we deal only with \emph{closed} queries i.e., closed
$\mathcal{L}$-formulas. The query answers are Boolean: $\true$ or
$\false$. A query is \emph{atomic} (\emph{quantifier-free}) if the
$\mathcal{L}$-formula is atomic (quantifier-free respectively). A {\em
  conjunctive} query is an existentially quantified conjunction of
atomic $\mathcal{L}$-formulas.
\begin{definition}
  Given an instance $I$ and a closed query $Q$, {\bf true} is the
  answer to $Q$ in $I$ if $I\models Q$; otherwise the answer to $Q$ in
  $I$ is $\false$.
\end{definition}

\subsection{Repairing}
In the original framework, when repairing a database two operations
are considered: inserting a fact and deleting a fact. In the presence
of denial constraints inserting facts cannot resolve inconsistencies,
and thus the repairs of the original instance are obtained by deleting
facts only i.e., the repairs are subsets of the original instance.
\begin{definition}[Repair]\label{def:repair}
  Given an instance $I$ and a set of denial constraints $F$, an
  instance $I'$ is a \emph{repair} of $I$ w.r.t.\ $F$ if and only if
  $I'$ is a maximal subset of $I$ that is consistent with $F$. By
  $\Rep(I,F)$ we denote the set of all repairs of $I$ w.r.t.\ $F$.
\end{definition}
To identify the facts whose mutual presence causes inconsistency we
use the notion of a conflict.
\begin{definition}[Conflict]
  Given a instance $I$ and a set of denial constraints $F$, a set of
  facts $\{R_1(t_1),\ldots,R_n(t_n)\}\subseteq I$ is a \emph{conflict}
  in $I$ w.r.t.\  $F$ if for some denial constraint in $F$ of the form
  \[
  \forall\bar{x} .\  \neg[R_1(\bar{x}_1)\land \ldots \land
  R_n(\bar{x}_n)\land\varphi(\bar{x})]
  \]
  there exists a substitution $\rho$ of variables $\bar{x}$ such that
  $\varphi(\rho(\bar{x}))$ is valid and $\rho(\bar{x}_i)=t_i$ for
  every $i\in\{1,\ldots,n\}$.
\end{definition}

We recall the notion of a conflict hypergraph that allows to visualize
all the conflicts present in the instance~\cite{ABCHRS03,ChMa04}.  We
recall that a hypergraph is a generalization of an undirected graph by
allowing more than two nodes to be connected by a hyperedge. Formally,
a \emph{hypergraph} is a pair consisting of a set of nodes and a set of
hyperedges, where a \emph{hyperedge} is a subset of the node set. Given
a hypergraph $\mathcal{G}$ we denote its set of nodes by
$V(\mathcal{G})$, and its set of hyperedges by $E(\mathcal{G})$.  
\begin{definition}[Conflict hypergraph]
Given a set of integrity constraints $F$ and a database instance $I$, the 
\emph{conflict hypergraph} $\mathcal{G}(I,F)$ of $I$ w.r.t.\ $F$ is a
hypergraph whose set of nodes is $I$ and set of hyperedges consists of
all conflicts in $I$ w.r.t.\ $F$.
\end{definition}
The size of the hypergraph is he sum of the size of the node set and the 
cardinalities of all hyperedges. We observe that assuming $F$ to be
fixed, the maximum cardinality of every hyperedge in a conflict
hypergraph is bounded from above by a constant. Consequently, the size 
of a conflict hypergraph $\mathcal{G}(I,F)$ is polynomial in the size of
the instance $I$.

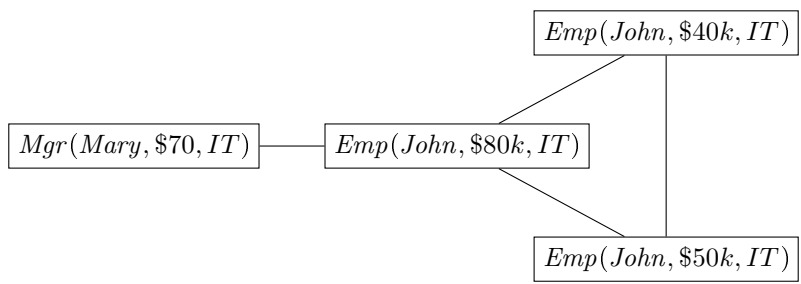
\begin{figure}[htb]
\begin{center}\small
\begin{tikzpicture}
\node[draw] at (0.5,0) (e1) {$\Emp(\John,\$80k,IT)$};
\node[draw] at (3.25,-1.5) (e2) {$\Emp(\John,\$50k,IT)$};
\node[draw] at (3.25, 1.5) (e3) {$\Emp(\John,\$40k,IT)$};
\node[draw] at (-3.75,0) (m1) {$\Mgr(\Mary,\$70,IT)$};

\draw[-] (m1) -- (e1);
\draw[-] (e1) -- (e2);
\draw[-] (e1) -- (e3);
\draw[-] (e2) -- (e3);
\end{tikzpicture}
\end{center}
\caption{\label{fig:conflict-hypergraph} The conflict graph $\mathcal{G}(I_0,F_0)$.}
\end{figure}

Two nodes are \emph{neighboring} (or are \emph{neighbors}) in a
hypergraph if there exists a hyperedge containing both nodes. The {\em
  neighborhood} of a node $v\in V(\mathcal{G})$ in a hypergraph
$\mathcal{G}$ is
\[
n_{\mathcal{G}}(v)=\{v'\in V(\mathcal{G}) \sep \exists e\in E(\mathcal{G}) .\  \{v,v'\}\subseteq e\}.
\]
A hyperedge connecting exactly two nodes is called simply an edge and
a hypergraph having only edges is called a graph. Similarly, we define
the conflict graph. The conflict graph for the instance in
Example~\ref{ex:repairs} is in Figure~\ref{fig:conflict-hypergraph}.
The conflict hypergraph is also a compact representation of all
repairs as we recall the following fact.
\begin{proposition}[\cite{ABCHRS03,ChMa04}]
  A \emph{maximal independent set} of $\mathcal{G}(I,F)$ is any maximal
  set of vertices that contains no hyperedge. Any maximal independent
  set is a repair of $I$ w.r.t.\ $F$ and vice versa.
\end{proposition}
\label{sec:clusters}
We recall that for only one key dependency (per relation name), the
conflict graph is a union of pairwise disjoint cliques and every
repair consists of exactly one element from each
clique~\cite{ABCHRS03}. To generalize this observation to FDs we
assume only one relation name $R$ and one functional dependency $R :
X\rightarrow Y$. Now, given an instance $I$, an \emph{$X$-cluster} is
the set of all facts (of $R$) in $I$ that have the same attribute
value in $X$, and similarly, an \emph{$(X,Y)$-cluster} is the set of
all facts (of $R$) in $I$ that have the same attribute value in $X$
and $Y$. Clearly, an $X$-cluster is a union of all $(X,Y)$-clusters
with the same attribute value in $X$. We recall that every repair
contains exactly one $(X,Y)$-cluster from each $X$-cluster. We also
remark that conflicts are present only inside an $X$-cluster and two
facts from the same $X$-cluster form a conflict if and only if they
belong to different $(X,Y)$-clusters.  
\begin{example}
  Consider the database schema consisting of exactly one relation name
  $R(A,B,C)$ and the FD $R:A\rightarrow B$. Take the following
  database instance
  \begin{multline*}
  I_2=\{R(1,1,1),R(1,1,2),R(1,1,3),R(1,2,1),R(1,2,2),\\
  R(2,1,1),R(2,1,2),R(2,1,3),R(2,2,1)\}.
  \end{multline*}
  Its conflict graph is presented in Figure~\ref{fig:clusters}. 
  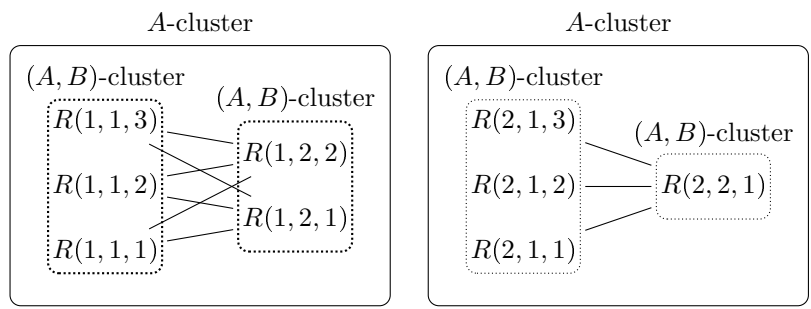
\begin{figure}[htb]
    \centering
    \begin{tikzpicture}[xscale=1,yscale=1.15]\small
      \node at (0,0.0) (n0) {$R(1,1,1)$};
      \node at (0,0.75)(n1) {$R(1,1,2)$};
      \node at (0,1.5) (n2) {$R(1,1,3)$};
      \node at (2.5,0.375) (n3) {$R(1,2,1)$};
      \node at (2.5,1.125) (n4) {$R(1,2,2)$};
      
      \draw (n0) -- (n3);
      \draw (n0) -- (n4);
      \draw (n1) -- (n3);
      \draw (n1) -- (n4);
      \draw (n2) -- (n3);
      \draw (n2) -- (n4);

      \node at (1.25,2.65) {$A$-cluster};
      \begin{scope}[yshift=0.125cm]
        \draw[rounded corners] 
        (-1.25,-0.75) -- (3.75,-0.75) -- (3.75,2.25) -- (-1.25,2.25) -- cycle;
      \end{scope}
      \node at (0,2) {$(A,B)$-cluster};
      \draw[rounded corners,densely dotted,thick] 
      (-0.75,-0.25) -- (0.75,-0.25) -- (0.75,1.75) -- (-0.75,1.75) -- cycle;
      \begin{scope}[xshift=2.5cm]
        \node at (0,1.75) {$(A,B)$-cluster};
        \draw[rounded corners,densely dotted,thick] 
        (-0.75,0) -- (0.75,0) -- (0.75,1.5) -- (-0.75,1.5) -- cycle;
      \end{scope}

      \node at (5.5,0.0) (n0)    {$R(2,1,1)$};
      \node at (5.5,0.75)(n1)    {$R(2,1,2)$};
      \node at (5.5,1.5) (n2)    {$R(2,1,3)$};
      \node at (8,0.75) (n3) {$R(2,2,1)$};

      \draw (n0) -- (n3);
      \draw (n1) -- (n3);
      \draw (n2) -- (n3);

      \begin{scope}[xshift=5.5cm]
      \node at (1.25,2.65) {$A$-cluster};
      \begin{scope}[yshift=0.125cm]
        \draw[rounded corners] 
        (-1.25,-0.75) -- (3.75,-0.75) -- (3.75,2.25) -- (-1.25,2.25) -- cycle;
      \end{scope}
      \node at (0,2) {$(A,B)$-cluster};
      \draw[rounded corners,densely dotted] 
      (-0.75,-0.25) -- (0.75,-0.25) -- (0.75,1.75) -- (-0.75,1.75) -- cycle;
      \begin{scope}[xshift=2.5cm]
        \node at (0,1.35) {$(A,B)$-cluster};
        \draw[rounded corners,densely dotted] 
        (-0.75,0.375) -- (0.75,0.375) -- (0.75,1.125) -- (-0.75,1.125) -- cycle;
      \end{scope}
      \end{scope}

    \end{tikzpicture}
    \caption{\label{fig:clusters}$A$- and $(A,B)$-clusters of $I_2$.}
  \end{figure}
  $I_2$ has two $A$-clusters each consisting of two $(A,B)$-clusters
  (indicated with a dotted line). For instance, the consider the
  $A$-cluster $\{R(2,1,1),R(2,1,2),R(2,1,3),R(2,2,1)\}$ which
  consists of two $(A,B)$-clusters: $\{R(2,2,1)\}$ and
  $\{R(2,1,1),R(2,1,2),R(2,1,3)\}$. \qed
\end{example}
Finally, we recall the basic database repairing
algorithm~\cite{StCh08}. 
\begin{algorithm}[htb]
\caption{\label{alg:repairing} Constructing a repair of $I$ w.r.t.\ $F$}
\begin{tabbing}
mm\=mm\=mm\=mm\=\kill
\step{1} \> $I^o \gets I$ \\
\step{2} \> $J\gets\varnothing$ \\
\step{3} \> {\bf while} $I^o\neq\varnothing$ {\bf do}\\
\step{4} \> \> {\bf choose} $R(t)\in I^o$\\
\step{5} \> \> $I^o\gets I^o\setminus\{R(t)\}$\\
\step{6} \> \> {\bf if } $J\cup\{R(t)\}\models F$ {\bf then}\\
\step{7} \> \> \> $J\gets J\cup\{R(t)\}$ \\
\step{8} \> {\bf return} $J$
\end{tabbing}
\vspace{-9pt}
\end{algorithm}
Algorithm~\ref{alg:repairing} iterates over the facts of the input
instance $I$ in some arbitrary order and creates a repair $J$. For
every fact it adds the fact to $J$ if so does not violate the set of
denial constraints $F$; otherwise the fact is discarded. Naturally,
the constructed instance $J$ is consistent with $F$. Moreover, $J$ is
a repair i.e., maximal consistent subset of $I$, since the algorithm
considers adding every fact to the constructed instance. Thus
Algorithm~\ref{alg:repairing} is \emph{sound}, it always produces a
repair.

We observe that depending on the order in which
Algorithm~\ref{alg:repairing} iterates over the facts in input
instance, we may obtain different repairs. For instance, in
Example~\ref{ex:repairs} the repair $I_2'$ is obtained with the
following ordering of the facts of $I_0$:
1) $\Mgr(\Mary,\$70k,IT)$, 
2) $\Emp(\John,\$50k,IT)$, 
3) $\Emp(\John,\$40k,IT)$, and 
4) $\Emp(\John,\$80k,IT)$.
On the other hand the repair $I_3'$ is obtained with the following
ordering of $I_0$:
1) $\Mgr(\Mary,\$70k,IT)$, 
2) $\Emp(\John,\$40k,IT)$, 
3) $\Emp(\John,\$50k,IT)$, and
4) $\Emp(\John,\$80k,IT)$.
In fact, for every repair $I'\in Rep(I,F)$ there exists an ordering of
$I$ for which Algorithm~\ref{alg:repairing} returns $I'$: it suffices
to take any ordering of $I'$ and append to it any ordering of
$I\setminus I'$. Hence, we say that Algorithm~\ref{alg:repairing} is
{\em{}complete} because it is capable of producing any repair.
\subsection{Complexity classes}
We make use of the following complexity classes:
\begin{itemize}
\item LOGSPACE: the class of decision problems solvable in logarithmic
  space by deterministic Turing machines (the input tape is read-only);
\item PTIME: the class of decision problems solvable in polynomial
  time by deterministic Turing machines;
\item coNP: the class of decision problems whose complements are
  solvable in polynomial time by nondeterministic Turing machines;
\item $\Pi^p_2$: the class of decision problems whose complements are
  solvable in polynomial time by nondeterministic Turing machines with
  an NP oracle.
\end{itemize}
We remark that these complexity classes are used only to measure the
\emph{data complexity} i.e., the complexity expressed in terms of the
size of the database size only~\cite{Var82}
(cf.\ Section~\ref{sec:data-complexity}).
\section{Conflict resolution preferences}
\label{sec:preferences}
To represent the preference information we use a relation on pairs of
neighboring facts i.e., pairs of facts present in a
conflict. Resolving a conflict consists of deleting one of its
elements and the relation is used to indicate those tuples that the
user prefers to keep in the database. We observe, however, that a
cycle in the relation may make the choice of the tuple to keep
ambiguous, if not impossible.  Consequently, we work with acyclic
relations only.
\begin{definition}[Priority]
  Given an instance $I$ and a set of denial constraints $F$, a {\em
    priority} $\succ$ of $I$ w.r.t.\ $F$ is a binary relation on $I$
  such that: (1) $\succ$ is acyclic and (2) for every $R(t),R'(t')\in
  I$ if $R(t) \succ R'(t')$, then $R(t)$ and $R'(t')$ are neighbors.
\end{definition}
In the sequel, we omit the reference to the instance $I$ and the set
of denial constraints $F$ if they are known from the context. 

From the point of the user interface it is often more natural to
define the priority as some acyclic binary relation on facts of $I$
and then consider the restriction of the priority relation to the
conflicting facts. Clearly, this approach can be handled with the
notion of priorities.

To help visualizing the priority we use the \emph{prioritized conflict
  hypergraphs}. Basically, we extend the conflict hypergraph with
directed edges corresponding to the priority relation:
$R(t)\rightarrow P(s)$ reads $R(t) \succ P(s)$.  The examples we
present in this paper use only conflict graphs i.e., conflict
hypergraphs where edges connect exactly two nodes. Consequently, a
prioritized graph can be seen as a graph with some of its edges
oriented. For instance,
Figure~\ref{fig:prioritized-conflict-hypergraph} contains the conflict
graph for the instance in Example~\ref{ex:repairs} with the priority
corresponding to the following preference: if the database contains an
employee who earns more than her manager, then the information about
the employee should be removed.
\begin{figure}[htb]
\begin{center}\small
\begin{tikzpicture}[>=stealth']
\node[draw] at (0.5,0) (e1) {$\Emp(\John,\$80k,IT)$};
\node[draw] at (3.25,-1.5) (e2) {$\Emp(\John,\$50k,IT)$};
\node[draw] at (3.25, 1.5) (e3) {$\Emp(\John,\$40k,IT)$};
\node[draw] at (-3.75,0) (m1) {$\Mgr(\Mary,\$70,IT)$};

\draw[->] (m1) -- (e1);
\draw[-] (e1) -- (e2);
\draw[-] (e1) -- (e3);
\draw[-] (e2) -- (e3);
\end{tikzpicture}
\caption{\label{fig:prioritized-conflict-hypergraph} Prioritized conflict graph.}
\end{center}
\end{figure}
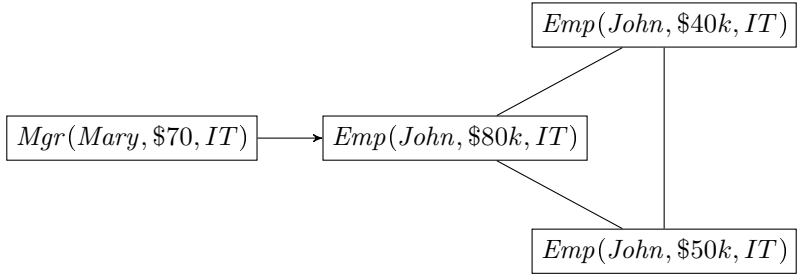
\begin{definition}[Priority extension]
  Given an instance $I$, a set of denial constraints $F$, and two
  priorities $\succ$ and $\succ'$ of $I$ w.r.t.\ $F$, $\succ'$ is an
  \emph{extension} of $\succ$, denoted $\mathord{\succ} \subseteq
  \mathord{\succ}'$ if and only if $R(t)\succ' R'(t')$ whenever
  $R(t)\succ R'(t')$ for $R(t),R'(t')\in I$. A priority $\succ$ of $I$
  w.r.t.\ $F$ is \emph{total} if there exists no priority $\succ'$ of
  $I$ w.r.t.\ $F$ that is different from $\succ$ and extends $\succ$.
\end{definition}
Note that both an extension of a priority and a total priority are
also acyclic and defined on pairs of neighboring facts only.
\begin{proposition}
  A priority $\succ$ is total if and only if for every conflict $C$
  and any two facts $x_1,x_2\in C$ we have that either $x_1\succ x_2$
  or $x_2\succ x_1$.
\end{proposition}
\begin{proof}
  The \emph{if} part is trivial. For the \emph{only if} part suppose
  there is a priority $\succ$ that is total yet there exists
  neighboring $x_1$ and $x_2$ such that $x_1\not\succ x_2$ and
  $x_2\not\succ x_1$ i.e., both
  $\mathord{\succ}_1=\mathord{\succ}\cup\{(x_1,x_2)\}$ and
  $\mathord{\succ}_2=\mathord{\succ}\cup\{(x_2,x_1)\}$ are
  cyclic. Since $\succ$ is not cyclic, $\succ_1$ has a cycle that
  traverses $(x_1,x_2)$ i.e., there exists a chain $x_2\succ y_1\succ
  \ldots \succ y_n\succ x_1$. Similarly, $\succ_2$ being cyclic
  implies that there exists a chain $x_1\succ z_1\succ \ldots \succ
  z_m \succ x_2$. Together this implies that $x_1 \succ\ldots\succ x_2
  \succ \ldots\succ x_1$; a contradiction. To finish the proof
  we observe that the acyclicity of priority implicitly excludes the
  possibility of both $x\succ y$ and $y\succ x$ being true at the same
  time for some facts $x$ and $y$.\qed
\end{proof}

\subsection{Preferred repairs and consistent query answers}
Now, we introduce the general framework of prioritized repairing and
query of inconsistent databases. We begin by defining a general notion
of a family of preferred repairs. We do not make any assumptions on
how such a family constructs preferred repairs. For generality, we do
not even assume that the constructed instances are repairs in the
sense of Definition~\ref{def:repair}. Instead, we list later on the
desirable properties that a well-behaved family should satisfy.

\begin{definition}[Preferred repairs]
  A \emph{family of preferred repairs} is a function $\XRep$
  defined on triplets $(I,F,\mathord{\succ})$, where $\succ$ is a
  priority in $I$ w.r.t.\ a set of denial constraints $F$, such that
  $\XRep(I,F,\mathord{\succ})$ is a set of database instances
  over the same schema. We say that a family $\YRep$ {\em
    subsumes} a family $\XRep$, denoted $\XRep
  \sqsubseteq \YRep$, if $\XRep(I,F,\mathord{\succ})
  \subseteq \YRep(I,F,\mathord{\succ})$ for every
  $(I,F,\mathord{\succ})$.
\end{definition}

We generalize the notion of consistent query answers ~\cite{ArBeCh99}
by considering only preferred repairs when evaluating a query (instead
of all repairs). We can easily generalize our approach to open queries
as in~\cite{ChMa04,ChMaSt04}.

\begin{definition}[$\mathcal{X}$-preferred consistent query answer]
\label{def:preferred-cqa}
Given a closed query $Q$, a triple $(I,F,\mathord{\succ})$, and a
family of preferred repairs $\XRep$, $\true$ ($\false$) is
the \emph{$\mathcal{X}$-preferred consistent query answer} to $Q$ in
$I$ w.r.t.\ $F$ and $\succ$ if for every
$I'\in\XRep(I,F,\mathord{\succ})$ we have $I' \models Q$
($I'\varnot\models Q$ respectively).
\end{definition}
Note that we obtain the original notion of consistent query answer if
we consider the family of all repairs $\Rep(I,F)$.

\subsection{Desirable properties of preferred repairs}
Now, we identify desirable properties of arbitrary families of
preferred repairs. The properties should be satisfied for an arbitrary
instance $I$ and an arbitrary set of denial constraints $F$.
\subsubsection*{$\mathcal{P}1$ {\bf Non-emptiness}}
Because the set of preferred repairs is used to define preferred
consistent query answers, it is important that for any preference
the framework is not trivialized by an empty set of
preferred repairs:
\begin{equation*}
  \XRep(I,F,\mathord\succ) \neq \varnothing. 
\end{equation*}

\subsubsection*{$\mathcal{P}2$ {\bf Monotonicity}}
The operation of extending the preference allows to improve the state
of our knowledge of the real world. The better such knowledge is the
finer the (preferred consistent) answers we should obtain. This is
achieved if extending the preference can only narrow the set of
preferred repairs:
\begin{equation*} 
  \mathord\succ_1 \subseteq \mathord\succ_2 
  \Longrightarrow
  \XRep(I,F,\mathord\succ_2) \subseteq 
  \XRep(I,F,\mathord\succ_1). 
\end{equation*}

\subsubsection*{$\mathcal{P}3$ {\bf Non-discrimination}}
Removing repairs from consideration must be justified by existing
preference. In particular, no repair should be removed if
no preference is given:
\begin{equation*}
  \XRep(I,F,\varnothing) = \Rep(I,F).
\end{equation*}

\subsubsection*{$\mathcal{P}4$ {\bf Categoricity}}
Ideally, a preference that cannot be further extended (the priority is
total) should specify how to resolve every conflict:
\begin{equation*}
  \text{$\succ$ is total} \Longrightarrow
  |\XRep(I,F,\mathord\succ)| = 1. 
\end{equation*}

\subsubsection*{$\mathcal{P}5$ {\bf Conservativeness}}
We also note that properties $\mathcal{P}2$ and $\mathcal{P}3$
together imply that preferred repairs are a subset of all repairs:
\begin{equation*}
  \XRep(I,F,\succ) \subseteq \Rep(I,F).
\end{equation*}
In fact, in the remainder of the paper we consider only families of
preferred repairs that satisfy $\mathcal{P}5$. We also observe that
$\mathcal{P}5$ with $\mathcal{P}1$ imply that the only preferred
repair of a consistent database instance is the instance itself.  
\subsection{Data complexity}
\label{sec:data-complexity}
We also adapt the decision problems to include the priority. Note that
the priority relation is of size quadratic in the size of the database
instance, and therefore, it is natural to make it a part of the
input. For a family $\XRep$ of preferred repairs the decision
problems we study are defined as follows:
\begin{enumerate}
\itemsep0pt
\item[$(i)$] \emph{$\mathcal{X}$-preferred repair checking} i.e., the
  complexity of the following set
\[
\mathcal{B}_{F}^\mathcal{X} = \{(I,\mathord{\succ},I') :
I'\in\XRep(I,F,\mathord{\succ})\}. 
\]
\item[$(ii) $] \emph{$\mathcal{X}$-preferred consistent query
  answering} i.e., the complexity of the following set  
\[
\mathcal{D}_{F,Q}^\mathcal{X} =
\{(I,\mathord{\succ}) : \forall I'\in\XRep(I,F,\mathord{\succ}). I'\models Q
\}.
\]
\end{enumerate}
\section{Globally-optimal repairs}
\label{sec:global}
We investigate several different families of preferred repairs.  The
first family of preferred repairs is based on the notion of optimal
compliance of the repair with the priority. Essentially, the
compliance of a repair can be improved by replacing a subset of facts
with a \emph{more preferred} subset of facts. The way we define a set
of facts being more preferred than another set of facts is inspired 
by the work on preferred models of logic programs~\cite{NiVe02} and
preferential reasoning~\cite{Ha97}.
\begin{definition}[Globally-optimal repairs $\GRep$]
\label{def:global}
Given an instance $I$, a set of denial constraints $F$, and a priority
$\succ$, an instance $I'\subseteq I$ is \emph{globally optimal}
w.r.t.\ $\succ$ and $F$ if no nonempty subset $X$ of facts from $I'$
can be replaced with a subset $Y$ of $I\setminus I'$ such that
\begin{equation}\tag{$\Asterisk_\mathcal{G}$}\label{eq:global-optimality}
\forall x \in X.\  \exists y \in Y.\  y \succ x
\end{equation}
and the resulting set of facts is consistent with
$F$. $\GRep$ is the family of globally-optimal repairs
i.e., $\GRep(I,F,\mathord{\succ})$ is the set of all repairs
of $I$ w.r.t.\ $F$ that are globally optimal w.r.t.\ $\succ$ and $F$.
\end{definition}
We emphasize that the family $\GRep$ selects all
globally-optimal repairs. In general, it is, however, possible to
define a family that selects only some of the globally-optimal
repairs, or even more generally, a family that constructs a set of
globally-optimal instances that need not be repairs.

The notion of global optimality identifies repairs whose compliance
with the priority cannot be further improved. For the instance $I_0$
in Example~\ref{ex:repairs} with the priority in
Figure~\ref{fig:prioritized-conflict-hypergraph} the set of 
globally-optimal repairs consists of $I_2'$ and $I_3'$.

In the sequel, we fix an instance $I$ and a set of denial constraints
$F$, and omit them when referring to the elements of
$\GRep(I,F,\mathord\succ)$. Before investigating the
properties of $\GRep$ we present an alternative
characterization of globally-optimal repairs.
\begin{proposition}\label{prop:global-alternative-definition}
  For a given priority $\succ$ and two repairs $I_1'$ and $I_2'$,
  $I_1'$ \emph{globally dominates} $I_2'$, denoted $I_1'
  \gg_\mathcal{G} I_2'$, if
\begin{equation}\tag{$\bigvarstar_\mathcal{G}$}\label{eq:global-domination}
\forall x \in I_2' \setminus I_1' . \; 
\exists y \in I_1' \setminus I_2' . \; y \succ x.
\end{equation}
The following facts hold:
\begin{enumerate}
\item[(i)] a repair $I'$ is globally optimal if and only if it is
  $\gg_\mathcal{G}$-maximal i.e., there is no repair $I''$ different
  from $I'$ such that $I''\gg_\mathcal{G} I'$;
\item[(ii)] if $\succ$ is acyclic, then so is $\gg_\mathcal{G}$.
\end{enumerate}
\end{proposition}
\begin{proof} {\it (i)} We prove the contraposition i.e., $I'$ is not
  globally optimal if and only if there exists a repair $I''\neq I'$
  such that $I''\gg_\mathcal{G} I'$. For the \emph{if}\/ part take
  $X=I'\setminus I''$ and $Y=I''\setminus I'$, and note that
  \eqref{eq:global-optimality} follows from
  \eqref{eq:global-domination}. Naturally, $(I'\setminus
  X)\cup{}Y=I''$ is consistent. For the \emph{only if} part take any
  nonempty $X\subseteq I'$ and $Y\subseteq I\setminus I'$ such that
  \eqref{eq:global-optimality} is satisfied and $J=(I'\setminus X)\cup
  Y$ is consistent. We take any repair $I''$ that contains $J$. Such a
  repair exists since $J$ is consistent. Clearly, $I'\setminus
  I''\subseteq X$ and also $Y\subseteq I''\setminus I'$. Hence
  \eqref{eq:global-domination} follows from
  \eqref{eq:global-optimality}. Consequently, $I'$ is not globally
  optimal.

  {\it (ii)} Suppose $\gg_\mathcal{G}$ is cyclic i.e., there exists a
  sequence of different repairs $I_0',\ldots,I_{n-1}'$ such that
  $I_i'\gg I_{i+1}'$ for $i\in\{0,\ldots,n-1\}$, where the $+$
  operator is interpreted modulo $n$. We show that $\succ$ is cyclic
  as well. We construct inductively infinite sequences of facts
  $y_1,y_2,\ldots$ and numbers $k_1,k_2,\ldots$ such that
  $y_{j+1}\succ y_j$ for $j\in \mathbb{N}$ and $y_j\not\in I_{k_j}'$
  and $y_j \in I_{k_j+1}'$ for $j\in\mathbb{N}$.

  For $j=1$ let $y_1$ be any element of $I_1'\setminus I_0'$ and
  $k_1=1$. Now, suppose we have constructed the two sequences up to
  their $j$-th elements $y_j$ and $k_j$ such that $y_j\not\in
  I_{k_j}'$ and $y_j\in I_{k_j+1}'$. If $y_j\in I_0'$, then $y_j$ must
  have been \emph{pushed out} somewhere between $I_0'$ and $I_{k_j}'$
  i.e., there exists $k_{j+1}\in\{0,\ldots,k_j-1\}$ such that $y_j\in
  I_{k_{j+1}}'$ and $y_j\not\in I_{k_{j+1}+1}'$. By
  $I_{k_{j+1}+1}'\gg_\mathcal{G} I_{k_{j+1}}'$ there exists an element
  $y_{j+1}\in I_{k_{j+1}+1}'\setminus I_{k_{j+1}}'$ such that
  $y_{j+1}\succ y_j$.  The case when $y_j\not\in I_0'$ is treated
  symmetrically: $y_j$ must have been pushed out somewhere between
  $I_{k_j+1}'$ and $I_n'=I_0'$.

  Clearly, $I$ has only a finite number of elements and thus any
  infinite $\succ$-chain must have a repetition, and consequently
  $\succ$ is cyclic.\qed
\end{proof}
\begin{proposition}\label{prop:global-properties}
  $\GRep$ satisfies the properties $\mathcal{P}1$-$\mathcal{P}4$.  
\end{proposition}
\begin{proof}
  We get $\mathcal{P}1$ by acyclicity of $\gg_\mathcal{G}$ and
  Proposition~\ref{prop:global-alternative-definition}. To show
  $\mathcal{P}2$ we observe that if a repair is globally optimal
  w.r.t.\ $\succ_2$, then it is globally optimal w.r.t.\ any $\succ_1$
  such that
  $\mathord{\succ_1}\subseteq\mathord{\succ_2}$. $\mathcal{P}3$
  follows directly from definition: to show that a repair is not
  globally optimal, $\succ$ needs to be nonempty.

  Showing $\mathcal{P}4$ requires a more elaborate argument. Take a
  total $\succ$. By $\mathcal{P}1$ there exists at least one
  globally-optimal repair. Suppose that there exist two different
  globally-optimal repairs $I_0'$ and $I_1'$.  In the remaining part
  of the proof for $i\geq 2$ we let $I_i'=I_{i\mod 2}'$. We show that
  $\succ$ is cyclic by creating an infinite chain $\ldots\succ
  x_1\succ x_0$ such that $x_i\in I_i'\setminus I_{i+1}'$ for every
  $i\in\mathbb{N}$. For $x_0$ we take any element from $I_0'\setminus
  I_1'$. Now, assuming that the sequence has been defined up to the
  $i$-th element $x_i$, we choose $x_{i+1}$ to be any element of
  $I_{i+1}'\setminus I_i'$ such that $x_i\succ x_{i+1}$. We show the
  existence of $x_{i+1}$ using global optimality of $I_{i+1}'$ as
  follows. First, we observe that the instance $I_{i+1}'\cup\{x_i\}$
  is inconsistent since $x_i\not\in I_{i+1}'$ and $I_{i+1}'$ is a
  repair i.e., a maximal consistent subset of $I$.  Let $C_1,\ldots,
  C_k$ be all conflicts present in $I_{i+1}'\cup\{x_i\}$. Clearly, for
  every $j\in\{1,\ldots,k\}$ the conflict $C_j$ contains a fact
  $z_j\not\in I_i'$ since $C_j\varnot\subseteq I_i'$ by the
  consistency of $I_i'$. Let $X=\{z_1,\ldots,z_k\}$ and
  $Y=\{x_i\}$. Naturally, $(I_{i+1}'\setminus X) \cup Y$ is
  consistent, and thus by global optimality of $I_{i+1}'$ there exists
  an element $x_{i+1}\in X$ such that $x_i\not\succ x_{i+1}$. But by
  totality of $\succ$ and the fact that every element of $X$ is a
  neighbor of $x_i$, we have that $x_{i+1}\succ x_i$. Clearly,
  $x_{i+1}\not\in I_i'$, and moreover, $x_{i+1}\in I_{i+1}'$ because
  $x_{i+1}\in C_j\setminus\{x_i\}\subseteq I_{i+1}'$ for some
  $j\in\{1,\ldots,k\}$. This shows that $\succ$ is cyclic; a
  contradiction. \qed
\end{proof}

Now, we present Algorithm~\ref{alg:global-repairing} that constructs
globally-optimal repairs. It begins with an arbitrary repair $I'$
obtained with Algorithm~\ref{alg:repairing} and then iteratively
attempts to improve the compliance of the repair with the priority. At 
each iteration it replaces a subset $X\subseteq I'$ of facts with a more
preferred subset $Y\subseteq I\setminus I'$ and extends the
obtained consistent instance $J=(I'\setminus X)\cup Y$ to a repair $I''$ in 
a manner analogous to the way Algorithm~\ref{alg:repairing} creates a 
repair: by attempting to add to $J$ any fact from $I\setminus J$ as 
long as doing so does not create a conflict. 

\begin{algorithm}[htb]
\caption{\label{alg:global-repairing} Constructing a globally-optimal repair of $I$ w.r.t.\ $F$}
\begin{tabbing}
mm\=mm\=mm\=mm\=\kill
\step{1} \> {\bf construct a repair} $I'$ \quad \emph{/*Algorithm~\ref{alg:repairing}*/}\\
\step{2} \> {\bf while} 
$\exists X \subseteq I'.\  \exists Y\subseteq I\setminus I'. 
\forall x\in X.\  \exists y\in Y.\  y\succ x$
{\bf do}\\
\step{3} \> \> $J\gets (I'\setminus X)\cup Y$\\ 
\step{4} \> \> {\bf extend} $J$ {\bf to a repair} $I''$ \quad\emph{/*Algorithm~\ref{alg:repairing}*/}\\
\step{7} \> \> $I'\gets I''$ \\
\step{8} \> {\bf return} $I'$
\end{tabbing}%
\vspace{-9pt}
\end{algorithm}
Naturally, Algorithm~\ref{alg:global-repairing} is sound because its
main loop stops only if the instance $I'$ is globally optimal and
since $\gg_\mathcal{G}$ is acyclic, the loop always terminates. It is
also complete because it is based on Algorithm~\ref{alg:repairing}
which constructs any repair, in particular any globally-optimal repair
can be constructed in the line~\step{1} of
Algorithm~\ref{alg:global-repairing}. We observe that if $I'_i$ is the
repair constructed in the $i$-th iteration of the main loop, then
$I'_{i+1}\gg_\mathcal{G} I'_i$. Since $\gg_\mathcal{G}$ is acyclic and
the number of repairs bounded by an exponential function of the size
of $I$, the algorithm performs at most an exponential number of
iterations. Checking global optimality (line~\step{2}) can be done in
exponential time, and thus the algorithm works in exponential time.
\begin{theorem}
  Algorithm~\ref{alg:global-repairing} is a sound and complete
  algorithm constructing globally-optimal repairs. It works in time
  exponential in the size of the input instance and the priority
  relation.
\end{theorem}

Algorithm~\ref{alg:global-repairing} follows a rather simple
principle: start with an arbitrary repair and iteratively improve its
compliance with the priority until an optimal one is obtained. For
such an approach to be tractable, two concerns would need to be
addressed: 1) the preferred repair checking problem needs to be in
PTIME and 2) the number of possible iterations needs to be bounded by
a polynomial. Later on we show that $\mathcal{G}$-preferred repair
checking is coNP-complete
(Theorem~\ref{thm:globally-optimal-cqa-intractable}) which shows that
this approach cannot be tractable (unless $\mathrm{P}=\mathrm{NP}$),
and furthermore, it suggests that there does not exist a tractable
sound and complete algorithm constructing $\mathcal{G}$-preferred
repairs. However, for other families of preferred repairs considered
in this paper the preferred repair checking problem is in PTIME. In
the following example we construct a $\ll_{\mathcal{G}}$-chain of
exponential length, thus showing that the number of iterations of
Algorithm~\ref{alg:global-repairing} may be exponential. The same
construction shows that for the other families of repairs an algorithm
based on the same principle might require an exponential number of
iterations. Consequently, more sophisticated solutions are required.
\begin{example}\label{ex:exp-chain}
For a given $n\in\mathbb{N}$ we construct an instance $I_n$ and a
priority $\succ_n$ such that the size of $I_n$ is $O(n)$, the size of
$\succ_n$ is $O(n^2)$, and there exists a $\gg_G$-chain of length
$\Omega(2^n)$.

Intuitively, we construct a chain of repairs which emulates a $n$-bit
binary counter, incremented from $0=(0\cdots 0)_2$ to $2^n-1=(1\cdots
1)_2$. Incrementing a counter consists of setting to $1$ the least
significant bit with value $0$ and setting to $0$ all the preceding
bits (up to this point all set to $1$). 
For instance, if $n=3$ and we wish to increment the number
$3=(011)_2$, then we obtain $4=(100)_2$ by setting to $1$ the third
bit and setting to $0$ the first and second bit. This operation can be
seen as a (cascading) propagation of the carry bit. Notice that even
numbers have their least significant bit set to $0$ and thus require
no propagation of the carry bit.

We work with instances of one relation only $R(A,B)$ and 
the constructed instance $I_n$ comprises of the following facts:
\begin{itemize}
\item $p_i^0=R(i,0)$ representing the $i$-th bit set to $0$, for
  $i\in\{0,\ldots,n-1\}$
\item $p_i^1=R(i,1)$ representing the $i$-th bit set to $1$, for
  $i\in\{0,\ldots,n-1\}$,
\item $p_i^c=R(i,2)$ representing the $i$-th bit being carried over to
  the $(i+1)$-th bit, for $i\in\{0,\ldots,n-2\}$.
\end{itemize}
To ensure proper behavior of the counter we use the following three constraints:
\begin{align*}
& R : A \rightarrow B,\\
& \forall i,j.\  \neg [R(i,2) \land R(j,1) \land i > j],\\
& \forall i,j.\  \neg [R(i,1) \land R(j,2) \land j=i-1].
\end{align*}
The first constraint ensures that a bit is set to $0$, set to $1$, or
being carried to the higher bit. The second constraint ensures that
propagating a carry bit resets all lower bits to $0$. The third
constraint ensures that a bit can be carried over only if the
immediately higher bit is set to $0$. The correct order of increment
is ensured by the priority relation $\succ_n$ defined as:
\begin{align*}
  & p_i^1 \succ_n p_i^0     &
  & \text{for $i\in\{0,\ldots,n-1\}$,}\\
  & p_i^1 \succ_n p_{i-1}^c &
  & \text{for $i\in\{1,\ldots,n-1\}$,}\\
  & p_i^c \succ_n p_j^1     &
  & \text{for $i\in\{1,\ldots,n-2\}$ and $j\in\{0,\ldots,i\}$.}
\end{align*}
Notice that $p_i^x \succ_n p_j^y$ implies that either $i>j$ or $i=j$, 
$x=1$, and $y=0$. Consequently, $\succ_n$ is acyclic.

Now, we construct a $\ll_\mathcal{G}$-chain of repairs that
corresponds to subsequent natural numbers ranging from $0$ to
$2^n-1$. Additionally, for odd numbers the chain contains also repairs
that represent the cascading propagation of the carry bit.
Figure~\ref{fig:chain} contains an example of an instance $I_3$ and a
sequence of repairs that constitutes a $\gg_\mathcal{G}$-chain.
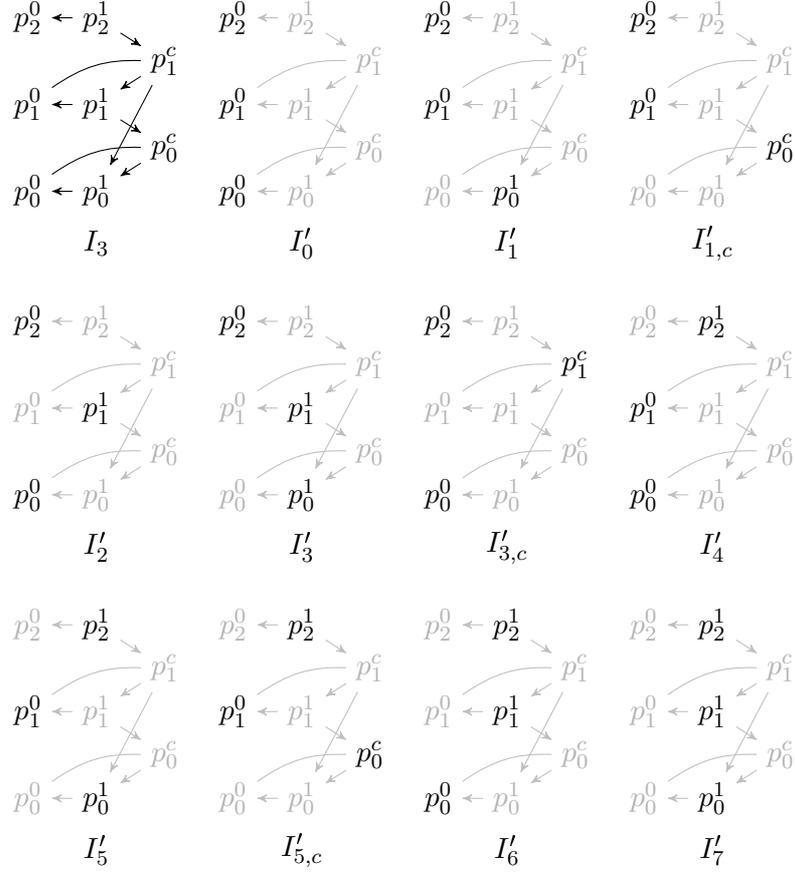
\begin{figure}[htb!]%
\begin{center}
\begin{tikzpicture}[>=stealth',yscale=1.15,xscale=0.9]
\tikzstyle{light}=[color=black!25!white, text=black!30!white]
\node at (0,0.0) (p00) {$p_0^0$};
\node at (1,0.0) (p01) {$p_0^1$};
\node at (2,0.5) (p0c) {$p_0^c$};

\node at (0,1.0) (p10) {$p_1^0$};
\node at (1,1.0) (p11) {$p_1^1$};
\node at (2,1.5) (p1c) {$p_1^c$};

\node at (0,2.0) (p20) {$p_2^0$};
\node at (1,2.0) (p21) {$p_2^1$};

\node at (1,-.6) (t) {$I_3$};

\draw[->] (p01) -- (p00);
\draw[<-] (p01) -- (p0c);
\draw[-, bend angle=15] (p00) edge[bend left] (p0c);

\draw[->] (p11) -- (p0c);

\draw[->] (p11) -- (p10);
\draw[<-] (p11) -- (p1c);
\draw[<-] (p01) -- (p1c);
\draw[-, bend angle=15] (p10) edge[bend left] (p1c);

\draw[->] (p21) -- (p1c);

\draw[->] (p21) -- (p20);

\begin{scope}[xshift=3cm]
\node at (0,0.0) (p00) {$p_0^0$};
\node[style=light] at (1,0.0) (p01) {$p_0^1$};
\node[style=light] at (2,0.5) (p0c) {$p_0^c$};

\node at (0,1.0) (p10) {$p_1^0$};
\node[style=light] at (1,1.0) (p11) {$p_1^1$};
\node[style=light] at (2,1.5) (p1c) {$p_1^c$};

\node at (0,2.0) (p20) {$p_2^0$};
\node[style=light] at (1,2.0) (p21) {$p_2^1$};

\node at (1,-.6) (t) {$I'_0$};

\draw[->][style=light] (p01) -- (p00);
\draw[<-][style=light] (p01) -- (p0c);
\draw[-, bend angle=15][style=light] (p00) edge[bend left] (p0c);

\draw[->][style=light] (p11) -- (p0c);

\draw[->][style=light] (p11) -- (p10);
\draw[<-][style=light] (p11) -- (p1c);
\draw[<-][style=light] (p01) -- (p1c);
\draw[-, bend angle=15][style=light] (p10) edge[bend left] (p1c);

\draw[->][style=light] (p21) -- (p1c);

\draw[->][style=light] (p21) -- (p20);
\end{scope}
\begin{scope}[xshift=6cm]
\node[style=light] at (0,0.0) (p00) {$p_0^0$};
\node at (1,0.0) (p01) {$p_0^1$};
\node[style=light] at (2,0.5) (p0c) {$p_0^c$};

\node at (0,1.0) (p10) {$p_1^0$};
\node[style=light] at (1,1.0) (p11) {$p_1^1$};
\node[style=light] at (2,1.5) (p1c) {$p_1^c$};

\node at (0,2.0) (p20) {$p_2^0$};
\node[style=light] at (1,2.0) (p21) {$p_2^1$};

\node at (1,-.6) (t) {$I'_1$};

\draw[->][style=light] (p01) -- (p00);
\draw[<-][style=light] (p01) -- (p0c);
\draw[-, bend angle=15][style=light] (p00) edge[bend left] (p0c);

\draw[->][style=light] (p11) -- (p0c);

\draw[->][style=light] (p11) -- (p10);
\draw[<-][style=light] (p11) -- (p1c);
\draw[<-][style=light] (p01) -- (p1c);
\draw[-, bend angle=15][style=light] (p10) edge[bend left] (p1c);

\draw[->][style=light] (p21) -- (p1c);

\draw[->][style=light] (p21) -- (p20);
\end{scope}
\begin{scope}[xshift=9cm]
\node[style=light] at (0,0.0) (p00) {$p_0^0$};
\node[style=light] at (1,0.0) (p01) {$p_0^1$};
\node at (2,0.5) (p0c) {$p_0^c$};

\node at (0,1.0) (p10) {$p_1^0$};
\node[style=light] at (1,1.0) (p11) {$p_1^1$};
\node[style=light] at (2,1.5) (p1c) {$p_1^c$};

\node at (0,2.0) (p20) {$p_2^0$};
\node[style=light] at (1,2.0) (p21) {$p_2^1$};

\node at (1,-.6) (t) {$I'_{1,c}$};

\draw[->][style=light] (p01) -- (p00);
\draw[<-][style=light] (p01) -- (p0c);
\draw[-, bend angle=15][style=light] (p00) edge[bend left] (p0c);

\draw[->][style=light] (p11) -- (p0c);

\draw[->][style=light] (p11) -- (p10);
\draw[<-][style=light] (p11) -- (p1c);
\draw[<-][style=light] (p01) -- (p1c);
\draw[-, bend angle=15][style=light] (p10) edge[bend left] (p1c);

\draw[->][style=light] (p21) -- (p1c);

\draw[->][style=light] (p21) -- (p20);
\end{scope}

\begin{scope}[yshift=-3.5cm]

\begin{scope}[xshift=0cm]
\node at (0,0.0) (p00) {$p_0^0$};
\node[style=light] at (1,0.0) (p01) {$p_0^1$};
\node[style=light] at (2,0.5) (p0c) {$p_0^c$};

\node[style=light] at (0,1.0) (p10) {$p_1^0$};
\node at (1,1.0) (p11) {$p_1^1$};
\node[style=light] at (2,1.5) (p1c) {$p_1^c$};

\node at (0,2.0) (p20) {$p_2^0$};
\node[style=light] at (1,2.0) (p21) {$p_2^1$};

\node at (1,-.6) (t) {$I'_2$};

\draw[->][style=light] (p01) -- (p00);
\draw[<-][style=light] (p01) -- (p0c);
\draw[-, bend angle=15][style=light] (p00) edge[bend left] (p0c);

\draw[->][style=light] (p11) -- (p0c);

\draw[->][style=light] (p11) -- (p10);
\draw[<-][style=light] (p11) -- (p1c);
\draw[<-][style=light] (p01) -- (p1c);
\draw[-, bend angle=15][style=light] (p10) edge[bend left] (p1c);

\draw[->][style=light] (p21) -- (p1c);

\draw[->][style=light] (p21) -- (p20);
\end{scope}
\begin{scope}[xshift=3cm]
\node[style=light] at (0,0.0) (p00) {$p_0^0$};
\node at (1,0.0) (p01) {$p_0^1$};
\node[style=light] at (2,0.5) (p0c) {$p_0^c$};

\node[style=light] at (0,1.0) (p10) {$p_1^0$};
\node at (1,1.0) (p11) {$p_1^1$};
\node[style=light] at (2,1.5) (p1c) {$p_1^c$};

\node at (0,2.0) (p20) {$p_2^0$};
\node[style=light] at (1,2.0) (p21) {$p_2^1$};

\node at (1,-.6) (t) {$I'_3$};

\draw[->][style=light] (p01) -- (p00);
\draw[<-][style=light] (p01) -- (p0c);
\draw[-, bend angle=15][style=light] (p00) edge[bend left] (p0c);

\draw[->][style=light] (p11) -- (p0c);

\draw[->][style=light] (p11) -- (p10);
\draw[<-][style=light] (p11) -- (p1c);
\draw[<-][style=light] (p01) -- (p1c);
\draw[-, bend angle=15][style=light] (p10) edge[bend left] (p1c);

\draw[->][style=light] (p21) -- (p1c);

\draw[->][style=light] (p21) -- (p20);
\end{scope}
\begin{scope}[xshift=6cm]
\node at (0,0.0) (p00) {$p_0^0$};
\node[style=light] at (1,0.0) (p01) {$p_0^1$};
\node[style=light] at (2,0.5) (p0c) {$p_0^c$};

\node[style=light] at (0,1.0) (p10) {$p_1^0$};
\node[style=light] at (1,1.0) (p11) {$p_1^1$};
\node at (2,1.5) (p1c) {$p_1^c$};

\node at (0,2.0) (p20) {$p_2^0$};
\node[style=light] at (1,2.0) (p21) {$p_2^1$};

\node at (1,-.6) (t) {$I'_{3,c}$};

\draw[->][style=light] (p01) -- (p00);
\draw[<-][style=light] (p01) -- (p0c);
\draw[-, bend angle=15][style=light] (p00) edge[bend left] (p0c);

\draw[->][style=light] (p11) -- (p0c);

\draw[->][style=light] (p11) -- (p10);
\draw[<-][style=light] (p11) -- (p1c);
\draw[<-][style=light] (p01) -- (p1c);
\draw[-, bend angle=15][style=light] (p10) edge[bend left] (p1c);

\draw[->][style=light] (p21) -- (p1c);

\draw[->][style=light] (p21) -- (p20);
\end{scope}
\begin{scope}[xshift=9cm]
\node at (0,0.0) (p00) {$p_0^0$};
\node[style=light] at (1,0.0) (p01) {$p_0^1$};
\node[style=light] at (2,0.5) (p0c) {$p_0^c$};

\node at (0,1.0) (p10) {$p_1^0$};
\node[style=light] at (1,1.0) (p11) {$p_1^1$};
\node[style=light] at (2,1.5) (p1c) {$p_1^c$};

\node[style=light] at (0,2.0) (p20) {$p_2^0$};
\node at (1,2.0) (p21) {$p_2^1$};

\node at (1,-.6) (t) {$I'_4$};

\draw[->][style=light] (p01) -- (p00);
\draw[<-][style=light] (p01) -- (p0c);
\draw[-, bend angle=15][style=light] (p00) edge[bend left] (p0c);

\draw[->][style=light] (p11) -- (p0c);

\draw[->][style=light] (p11) -- (p10);
\draw[<-][style=light] (p11) -- (p1c);
\draw[<-][style=light] (p01) -- (p1c);
\draw[-, bend angle=15][style=light] (p10) edge[bend left] (p1c);

\draw[->][style=light] (p21) -- (p1c);

\draw[->][style=light] (p21) -- (p20);
\end{scope}
\end{scope}
\begin{scope}[yshift=-7cm]

\begin{scope}[xshift=0cm]
\node[style=light] at (0,0.0) (p00) {$p_0^0$};
\node at (1,0.0) (p01) {$p_0^1$};
\node[style=light] at (2,0.5) (p0c) {$p_0^c$};

\node at (0,1.0) (p10) {$p_1^0$};
\node[style=light] at (1,1.0) (p11) {$p_1^1$};
\node[style=light] at (2,1.5) (p1c) {$p_1^c$};

\node[style=light] at (0,2.0) (p20) {$p_2^0$};
\node at (1,2.0) (p21) {$p_2^1$};

\node at (1,-.6) (t) {$I'_{5}$};

\draw[->][style=light] (p01) -- (p00);
\draw[<-][style=light] (p01) -- (p0c);
\draw[-, bend angle=15][style=light] (p00) edge[bend left] (p0c);

\draw[->][style=light] (p11) -- (p0c);

\draw[->][style=light] (p11) -- (p10);
\draw[<-][style=light] (p11) -- (p1c);
\draw[<-][style=light] (p01) -- (p1c);
\draw[-, bend angle=15][style=light] (p10) edge[bend left] (p1c);

\draw[->][style=light] (p21) -- (p1c);

\draw[->][style=light] (p21) -- (p20);
\end{scope}
\begin{scope}[xshift=3cm]
\node[style=light] at (0,0.0) (p00) {$p_0^0$};
\node[style=light] at (1,0.0) (p01) {$p_0^1$};
\node at (2,0.5) (p0c) {$p_0^c$};

\node at (0,1.0) (p10) {$p_1^0$};
\node[style=light] at (1,1.0) (p11) {$p_1^1$};
\node[style=light] at (2,1.5) (p1c) {$p_1^c$};

\node[style=light] at (0,2.0) (p20) {$p_2^0$};
\node at (1,2.0) (p21) {$p_2^1$};

\node at (1,-.6) (t) {$I'_{5,c}$};

\draw[->][style=light] (p01) -- (p00);
\draw[<-][style=light] (p01) -- (p0c);
\draw[-, bend angle=15][style=light] (p00) edge[bend left] (p0c);

\draw[->][style=light] (p11) -- (p0c);

\draw[->][style=light] (p11) -- (p10);
\draw[<-][style=light] (p11) -- (p1c);
\draw[<-][style=light] (p01) -- (p1c);
\draw[-, bend angle=15][style=light] (p10) edge[bend left] (p1c);

\draw[->][style=light] (p21) -- (p1c);

\draw[->][style=light] (p21) -- (p20);
\end{scope}
\begin{scope}[xshift=6cm]
\node at (0,0.0) (p00) {$p_0^0$};
\node[style=light] at (1,0.0) (p01) {$p_0^1$};
\node[style=light] at (2,0.5) (p0c) {$p_0^c$};

\node[style=light] at (0,1.0) (p10) {$p_1^0$};
\node at (1,1.0) (p11) {$p_1^1$};
\node[style=light] at (2,1.5) (p1c) {$p_1^c$};

\node[style=light] at (0,2.0) (p20) {$p_2^0$};
\node at (1,2.0) (p21) {$p_2^1$};

\node at (1,-.6) (t) {$I'_6$};

\draw[->][style=light] (p01) -- (p00);
\draw[<-][style=light] (p01) -- (p0c);
\draw[-, bend angle=15][style=light] (p00) edge[bend left] (p0c);

\draw[->][style=light] (p11) -- (p0c);

\draw[->][style=light] (p11) -- (p10);
\draw[<-][style=light] (p11) -- (p1c);
\draw[<-][style=light] (p01) -- (p1c);
\draw[-, bend angle=15][style=light] (p10) edge[bend left] (p1c);

\draw[->][style=light] (p21) -- (p1c);

\draw[->][style=light] (p21) -- (p20);
\end{scope}
\begin{scope}[xshift=9cm]
\node[style=light] at (0,0.0) (p00) {$p_0^0$};
\node at (1,0.0) (p01) {$p_0^1$};
\node[style=light] at (2,0.5) (p0c) {$p_0^c$};

\node[style=light] at (0,1.0) (p10) {$p_1^0$};
\node at (1,1.0) (p11) {$p_1^1$};
\node[style=light] at (2,1.5) (p1c) {$p_1^c$};

\node[style=light] at (0,2.0) (p20) {$p_2^0$};
\node at (1,2.0) (p21) {$p_2^1$};

\node at (1,-.6) (t) {$I'_7$};

\draw[->][style=light] (p01) -- (p00);
\draw[<-][style=light] (p01) -- (p0c);
\draw[-, bend angle=15][style=light] (p00) edge[bend left] (p0c);

\draw[->][style=light] (p11) -- (p0c);

\draw[->][style=light] (p11) -- (p10);
\draw[<-][style=light] (p11) -- (p1c);
\draw[<-][style=light] (p01) -- (p1c);
\draw[-, bend angle=15][style=light] (p10) edge[bend left] (p1c);

\draw[->][style=light] (p21) -- (p1c);

\draw[->][style=light] (p21) -- (p20);
\end{scope}
\end{scope}
\end{tikzpicture}
\end{center}
\caption{\label{fig:chain} The instance $I_3$ and the chain 
$
I_7'\gg_\mathcal{G}I_6'\gg_\mathcal{G}I_{5,c}'\gg_\mathcal{G}
I_5'\gg_\mathcal{G}I_4'\gg_\mathcal{G}I_{3,c}'\gg_\mathcal{G}
I_3'\gg_\mathcal{G}I_2'\gg_\mathcal{G}I_{1,c}'\gg_\mathcal{G}
I_1'\gg_\mathcal{G}I_0'$.}
\end{figure}%

For instance, $I_0'$ and $I_1'$ correspond to $0=(000)_2$ and
$1=(001)_2$ respectively while $I_{1,c}'$ corresponds to $1$ being
incremented with a carry bit. For every $i\in\{0,\ldots,2^n-1\}$ let
$(b_0^i,b_1^i,\ldots,b_{n-1}^i)$ be the binary representation of $i$,
where $b_j^i\in\{0,1\}$ and $b_0^i$ denotes the least significant bit
i.e., $\sum_{j=0}^{n-1}2^jb_j^i = i$. The repair corresponding to
$i\in\{0,\ldots,2^n-1\}$ is
\[
I_i'=\{p_0^{b_0^i},p_1^{b_1^i},\ldots,p_{n-1}^{b_{n-1}^i}\}.
\]

For every odd $i\in\{1,3,\ldots,2^n-3\}$ we also construct the repair
that propagates the carry bit in a cascading fashion
\[
I_{i,c}'=\{p_0^0,\ldots,p_{j_i-2}^0,p_{j_i-1}^c,p_{j_i}^{b_{j_i}^i},\ldots,p_{n-1}^{b_{n-1}^i}\},
\]
where $j_i$ is the position of the least significant bit of the binary
representation of $i$ that is set to $0$ i.e., the minimal $j$ such
that $b_j^i=0$.  It can be easily shown that
\begin{multline*}
I_{2^n-1}'\gg_\mathcal{G}I_{2^n-2}'\gg_\mathcal{G}I_{2^n-3,c}'\gg_\mathcal{G}
I_{2^n-3,c}'\gg_\mathcal{G}\ldots\\
\ldots
\gg_\mathcal{G}I_{3,c}'\gg_\mathcal{G}
\gg_\mathcal{G}I_3'\gg_\mathcal{G}
\gg_\mathcal{G}I_2'\gg_\mathcal{G}
\gg_\mathcal{G}I_{1,c}'\gg_\mathcal{G}
I_1'\gg_\mathcal{G}I_0'. 
\end{multline*}

Finally, we observe that in the worst case scenario
Algorithm~\ref{alg:global-repairing} may traverse the full length of
the constructed chain during its execution with $I_n$ and
$\succ_n$. We remark, however, that in this example the
globally-optimal repair $I'_{2^n-1}$ may be attained in just one
iteration of the main loop i.e., $I'_{2^n-1}\gg_\mathcal{G} I_i'$ for
$i\in\{0,\ldots,2^n-2\}$ and $I_{2^n-1}'\gg_\mathcal{G} I_{i,c}'$ for
$i\in\{1,3,\ldots,2^n-3\}$. \qed
\end{example}

Now, we investigate computational properties of globally-optimal
repairs. We observe that verifying whether a repair $I'$ is not
globally optimal can be easily accomplished with a nondeterministic
Turing machine: it suffices to guess the sets $X$ and $Y$, verify that
$(I'\setminus X)\cup Y$ is consistent, and check that
\eqref{eq:global-optimality} holds. Consequently,
$\mathcal{B}^\mathcal{G}_F$ is in coNP. The membership of
$\mathcal{D}_{F,Q}^\mathcal{G}$ in $\Pi_2^p$ follows from
Definition~\ref{def:preferred-cqa}: $\true$ is not the
$\mathcal{G}$-preferred consistent answer to a query if the query is
not $\true$ in some globally-optimal repair.
\begin{proposition}
  $\mathcal{G}$-preferred repair checking is in coNP and
  $\mathcal{G}$-preferred consistent query answering is in $\Pi^p_2$.
\end{proposition}
The upper bounds are tight.
\begin{theorem}\label{thm:globally-optimal-cqa-intractable}
  There exists a set of 4 FDs and an atomic query for which
  $\mathcal{G}$-preferred repair checking is coNP-hard and
  $\mathcal{G}$-preferred consistent query answering is
  $\Pi^p_2$-hard.
\end{theorem}
\begin{proof}
  We show $\Pi^p_2$-hardness of $\mathcal{D}_{F,Q}^\mathcal{G}$ by
  reducing the satisfaction of $\forall^*\exists^*$QBF formulas to
  $\mathcal{D}_{F,Q}^\mathcal{G}$. Consider the following formula:
  \[
  \Psi = \forall x_1,\ldots{},x_n . \exists
  x_{n+1},\ldots{},x_{n+m}. \Phi,
  \]
  where $\Phi$ is (quantifier-free) 3CNF i.e., $\Phi$ equals to $c_1
  \land \ldots{} \land c_s$, and $c_k$ is a clause of three literals
  $\ell_{k,1} \lor \ell_{k,2} \lor \ell_{k,3}$ for
  $k\in\{1,\ldots,s\}$. We call the variables $x_1,\ldots,x_n$ {\em
    universal} and $x_{n+1},\ldots,x_{n+m}$ \emph{existential}. We use
  the function $q$ to identify the type of a variable with a given
  index: $q(i)=1$ for $i\leq n$ and $q(i)=0$ for $i>n$. We also use
  the following two auxiliary functions $var$ and $sgn$ on literals of
  $\Phi$:
  \begin{align*}
    & var(x_i) = var(\neg x_i) = i, & &sgn(x_i) = 1,& &sgn(\neg x_i) =
    -1.
  \end{align*}
  A \emph{valuation} is a (possibly partial) function assigning a
  Boolean value to the variables.

  We construct instances over the schema consisting of a single
  relation \[R(A_1, B_1, A_2, B_2, A_3, B_3, A_4, B_4).\] The set of
  integrity constraints is
  \[
  F=\{A_1\rightarrow B_1,A_2\rightarrow B_2,A_3\rightarrow
  B_3,A_4\rightarrow B_4\}.
  \]
  The constructed database instance $I_\Psi$ consists of the following
  facts:
  \begin{itemize}
  \item $v_i$ and $\bar{v}_i$ corresponding to the positive and
    negative valuations of $x_i$ resp. (for $i\in\{1,\ldots,n+m\}$)
    \begin{align*}
      &v_i=R(0,q(i),i,1,i,1,i,1),&
      &\bar{v}_i=R(0,q(i),i,-1,i,-1,i,-1),
    \end{align*}
  \item $d_k$ corresponding to the clause $c_k$ (for $k\in\{1,\ldots,s\}$)
    \[
    d_k = R(0,1,var(\ell_{k,1}),sgn(\ell_{k,1}),
    var(\ell_{k,2}),sgn(\ell_{k,2}), var(\ell_{k,3}),sgn(\ell_{k,3})
    ),
    \]
  \item $p_\exists$ and $p_\forall$ used to partition the set of all
    repairs into repairs that correspond to the valuations of
      existential and universal variables respectively:
    \[
    p_\exists= R(0,0,0,0,0,0,0,0), \qquad p_\forall =
    R(0,1,0,0,0,0,0,0).
    \]
  \end{itemize}
  For the ease of reference by $L_{k,p}$ we denote the fact
  corresponding to the satisfying valuation of literal $\ell_{k,p}$
  i.e.:
  \[
  L_{k,p} =
  \begin{cases}
    v_i & \text{when $\ell_{k,p}=x_i$,}\\
    \bar{v}_i & \text{when $\ell_{k,p}=\neg x_i$.}
  \end{cases}
  \]
  The constructed priority relation $\succ_\Psi$ is the minimal
  priority of $I_\Psi$ w.r.t.\ $F$ such that:
  \begin{align*}
    & v_i \succ_\Psi d_k, &
    & \text{if $c_k$ uses a positive literal $x_i$},\\
    & \bar{v}_i \succ_\Psi d_k, &
    & \text{if $c_k$ uses a negative literal $\neg x_i$},\\
    & p_\exists \succ_\Psi v_i, &
    & \text{for all $i\in\{1,\ldots{},n\}$},\\
    & p_\exists \succ_\Psi \bar{v}_i, &
    & \text{for all $i\in\{1,\ldots{},n\}$},\\
    & p_\exists \succ_\Psi p_\forall.
  \end{align*}
  Figure~\ref{fig:global-reduction} contains a prioritized conflict
  graph of the instance and the priority obtained for the formula:
  \[
  \Psi_0 = \forall x_1, x_2, x_3.\exists x_4, x_5.  (\neg x_1 \lor x_4
  \lor x_2)\land (\neg x_2 \lor \neg x_5 \lor \neg x_3).
  \]
  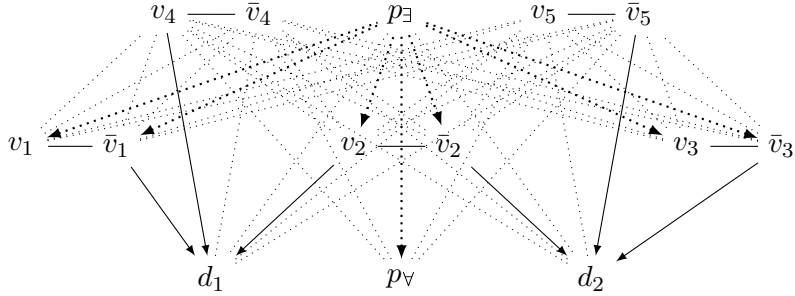
\begin{figure}[htb]
    \begin{center}
      \begin{tikzpicture}[xscale=1.25,yscale=1.75]
        \path[use as bounding box] (1,0) rectangle (9,2.25);
  
        \path (0,2) ++(2.5,0) node (q1) {$v_4$} ++(1,0) node (nq1)
        {$\bar{v}_4$} ++(1.5,0) node (y) {$p_\exists$} ++(1.5,0) node
        (q2) {$v_5$} ++(1,0) node (nq2) {$\bar{v}_5$};

        \path (0,1) ++(1,0) node[circle, inner sep=1mm] (p1) {$v_1$}
        ++(1,0) node (np1) {$\bar{v}_1$} ++(2.5,0) node (p2) {$v_2$}
        ++(1,0) node (np2) {$\bar{v}_2$} ++(2.5,0) node (p3) {$v_3$}
        ++(1,0) node (np3) {$\bar{v}_3$};

        \path (0,0) ++(3,0) node (d1) {$d_1$} ++(2,0) node (x)
        {$p_\forall$} ++(2,0) node (d2) {$d_2$};

        \draw (q1) -- (nq1) (q2) -- (nq2) (p1) -- (np1) (p2) -- (np2)
        (p3) -- (np3);

        \draw[dotted] (p1) -- (q1) (p1) -- (nq1) (p1) -- (q2) (p1) --
        (nq2) (np1) -- (q1) (np1) -- (nq1) (np1) -- (q2) (np1) --
        (nq2) (p2) -- (q1) (p2) -- (nq1) (p2) -- (q2) (p2) -- (nq2)
        (np2) -- (q1) (np2) -- (nq1) (np2) -- (q2) (np2) -- (nq2) (p3)
        -- (q1) (p3) -- (nq1) (p3) -- (q2) (p3) -- (nq2) (np3) -- (q1)
        (np3) -- (nq1) (np3) -- (q2) (np3) -- (nq2) (x) -- (q1) (x) --
        (nq1) (x) -- (q2) (x) -- (nq2) (d1) -- (nq1) (d1) -- (q2) (d1)
        -- (nq2) (d2) -- (q1) (d2) -- (nq1) (d2) -- (q2) (y) -- (d1)
        (y) -- (d2) ;

        \draw[latex-] (d1) -- (np1); \draw[latex-] (d1) -- (q1);
        \draw[latex-] (d1) -- (p2); \draw[latex-] (d2) -- (np2);
        \draw[latex-] (d2) -- (nq2); \draw[latex-] (d2) -- (np3);
        \draw[latex-,dotted, thick] (p1) -- (y); \draw[latex-,dotted,
        thick] (np1) -- (y); \draw[latex-,dotted, thick] (p2) -- (y);
        \draw[latex-,dotted, thick] (np2) -- (y); \draw[latex-,dotted,
        thick] (p3) -- (y); \draw[latex-,dotted, thick] (np3) -- (y);
        \draw[latex-,dotted, thick] (x) -- (y);
      \end{tikzpicture}
    \end{center}
    \caption{\label{fig:global-reduction} The prioritized conflict
      graph for $\Psi_0$. Dotted lines used for conflicts
      w.r.t.\ $A_1\rightarrow B_1$.  }
  \end{figure}

  The query used in the reduction is $Q=p_\exists$ and we claim that
  $\Psi$ is valid if and only if $\true$ is $\mathcal{G}$-preferred
  consistent query answer to $p_\exists$ in $I_\Psi$ w.r.t.\ $F$ and
  $\succ_\Psi$.  The proof is technically elaborate but can be
  summarized as follows. First, we partition the set of repairs into
  $\exists$- and $\forall$-repairs that correspond to valuations of
  existential and universal variables. Next, we show that an
  $\exists$-repair globally dominates a $\forall$-repair iff the
  combined valuation satisfies $\Phi$. Consequently, we argue that if
  the $\exists$-repairs are the only globally-optimal repairs, then
  for every valuation of universal variables there exists a valuation
  of existential variables that together satisfy $\Phi$ i.e., $\Psi$
  is valid.

  We partition the set of all repairs of $I_\Psi$ into two disjoint
  classes: {\it $\exists$-repairs} that contain $p_\exists$ and {\it
    $\forall$-repairs} that do not contain $p_\exists$. Because of the
  FD $A_1\rightarrow B_1$ every $\forall$-repair contains
  $p_\forall$. For the same reason, a $\forall$-repair is always a
  subset of
  $\{v_1,\bar{v}_1,\ldots,v_n,\bar{v}_n,d_1,\ldots,d_n,p_\forall\}$
  whereas an $\exists$-repair is always a subset of
  $\{v_{n+1},\bar{v}_{n+1},\ldots,v_{n+m},\bar{v}_{n+m},p_\exists\}$.

  We use $\exists$- and $\forall$-repairs to represent all possible
  valuation of existential and universal variables respectively. To
  easily move from a valuation of variables to a repair we define the
  following two operators:
  \begin{small}
    \begin{align*}
      I_\exists[V] ={} &\{v_i \sep V(x_i)=\true \land q(i)=0 \} \cup
      \{\bar{v}_i \sep V(x_i)=\false \land q(i)=0\} \cup \{p_\exists\},\\
      I_\forall[V] ={}& \{v_i \sep V(x_i)=\true \land q(i)=1\} \cup
      \{\bar{v}_i \sep V(x_i)=\false \land q(i)=1\} \cup \{p_\forall\} \cup{}\\
      &\left\{d_k \left|\, \text{\parbox{210pt}{ if for every literal
              $\ell_{k,i}$ of $c_k$ that uses a universal variable for
              which $V$ is defined, we have $V\varnot\models
              \ell_{k,i}$}} \right.\right\}.
    \end{align*}%
  \end{small}%
  Note that $I_\forall[V]$ contains the types corresponding to clauses
  that are not satisfied by the valuation of universal variables $V$
  alone.

  For instance, take the formula $\Psi_0$ in
  Figure~\ref{fig:global-reduction} and the following total
  valuation $V_0$ of variables $x_1,\ldots,x_5$:
  \begin{align*}
    &V_0(x_1) = \true,& 
    &V_0(x_2) = \false,&
    &V_0(x_3) = \false,&
    &V_0(x_4) = \true,&
    &V_0(x_5) = \true.
  \end{align*}
  Then, the repairs corresponding to the valuation of existential and
  universal variables are
  \[
  I_\exists[V_0]=\{v_4,v_5,p_\exists\}\quad\text{and}\quad
  I_\forall[V_0]=\{v_1,\bar{v}_2,\bar{v}_3,d_1,p_\forall\}.
  \]
  To move in the opposite direction, from a repair to a (possibly
  partial) valuation we use:
  \[
  V[I'](x_i) =
  \begin{cases}
    \true  &\text{if $v_i\in I'$},\\
    \false &\text{if $\bar{v}_i \in I'$},\\
    \mathrm{undefined} &\text{otherwise}.
  \end{cases}
  \]

  We observe that $V[\cdot]$ defines a one-to-one correspondence
  between $\exists$-repairs and total valuations of existential
  variables. A similar statement, however, does not hold for
  $\forall$-repairs because of the interaction between facts $d_k$ and
  the facts corresponding to universal variables. For example, for the
  instance in Figure~\ref{fig:global-reduction} if we take the repair
  $I_0=\{v_1,v_3,d_1,d_2,p_\forall\}$, the corresponding valuation
  $V[I_0]$ of universal variables is undefined for $x_2$.

  Consequently, for some $\forall$-repair $I'$ the function $V[I']$
  may be only a partial valuation of universal variables. We call a
  $\forall$-repair $I'$ \emph{strict} if $V[I']$ is a total valuation
  of universal variables. In this way, $V[\cdot]$ defines a one-to-one
  correspondence between strict $\forall$-repairs and total valuations
  of the universal variables. The following result allows us to remove
  non-strict $\forall$-repairs from consideration.
  \begin{lemma}\label{lemma:strict-are-maximal}
    Strict $\forall$-repairs are exactly $\gg_\mathcal{G}$-maximal
    $\forall$-repairs.
  \end{lemma}
  \begin{proof}
    First, we prove that no non-strict $\forall$-repair is
    $\gg_\mathcal{G}$-maximal. For that we show how to construct from
    a non-strict $\forall$-repair $I'$ a strict $\forall$-repair $I''$
    such that $I''\gg_\mathcal{G} I'$. We take the partial valuation
    $V'=V[I']$ and extend it to a total valuation $V''$ of universal
    variables by assigning $\false$ value to variables undefined by
    $V'$ i.e.,
    \[
    V'' = V' \cup \{(x_i,\false) \sep 1 \leq i \leq n \land
    \text{$V'(x_i)$ is undefined} \}.
    \]
    Now, we go back to the repair $I''=I_\forall[V'']$ and show that
    \[
    \forall q' \in I'\setminus I'' .\  \exists q'' \in I'' \setminus
    I'.\  q'' \succ q'.
    \]
    There are 4 cases of values of $q'$ to consider:
    \begin{enumerate}
    \item $q'=p_\exists$, $q'=v_i$, or $q'=\bar{v}_i$ for
      $i\in\{n+1,\ldots,n+m\}$ is not possible because neither of $I'$
      and $I''$ contains these facts (being $\forall$-repairs)
    \item $q'=p_\forall$ is not possible because both $I'$ and $I''$
      are $\forall$-repairs.
    \item $q'=v_i$ or $q'=\bar{v}_i$ for some $i\in\{1,\ldots,n\}$ is
      also impossible because from the construction of $I''$ we know
      that
      \[
      I'' \cap \{v_1,\bar{v}_1,\ldots,v_n,\bar{v}_n\} \subseteq I'
      \cap \{v_1,\bar{v}_1,\ldots,v_n,\bar{v}_n\}.
      \]
    \item $q'=d_k$ for some $k\in\{1,\ldots,s\}$. The neighborhood of
      $d_k$ in the conflict graph consists of facts $p_\exists$,
      $L_{k,1}$, $L_{k,2}$, and $L_{k,3}$. We observe that none of
      these facts belongs to $I'$. However, one of the facts must
      belong to $I''$ because $q'\not\in I''$ and $I''$ is a maximal
      consistent subset of $I_\Psi$. Since $I''$ is a
      $\forall$-repair, $p_\exists$ does not belong to
      $I''$. Therefore, for some $p\in\{1,2,3\}$ the fact $L_{k,p}$
      must belong to $I''$. Consequently, $q''=L_{k,p}\succ_\Psi q'$.
    \end{enumerate}

    Now, we show that every strict $\forall$-repair is also
    $\gg_\mathcal{G}$-maximal among $\forall$-repairs. Suppose
    otherwise i.e., for some strict $\forall$-repair $I'$ there exists
    an $\forall$-repair $I''$ such that $I'\gg_\mathcal{G} I''$. Since
    $I'$ is strict it contains $v_i$ or $\bar{v}_i$ for every
    $i\in\{1,\ldots,n\}$. By the construction of the priority
    $\succ_\Psi$ the repairs $I'$ and $I''$ must agree on facts
    $v_1,\bar{v}_1,\ldots,v_n,\bar{v}_n$. Therefore
    $I'=I_\forall[V[I'']]$ and using the reasoning from the previous
    part we can show that $I''\gg_\mathcal{G} I'$. Since $\succ_\Psi$
    is acyclic, by
    Proposition~\ref{prop:global-alternative-definition} this gives us
    $I'=I''$. \qed
  \end{proof}
  The central result in our reduction follows.
  \begin{lemma}\label{lemma:valuation-is-satisfaction}
    For any total valuation $V$, $I_\exists[V] \gg_\mathcal{G}
    I_\forall[V]$ if and only if $V\models \Phi$.
  \end{lemma}
  \begin{proof}
    For the \emph{if} part, because a $\forall$-repair is disjoint with
    any $\exists$-repair, it is enough to show that for any fact
    $q'\in I_\forall[V]$ there exists a fact $q'' \in I_\exists[V]$
    such that $q''\succ q'$. For
    $p_\forall,v_1,\bar{v}_1,\ldots,v_n,\bar{v}_n$ we simply choose
    $p_\exists$. If $d_k$ belongs to $I_\forall[V]$, then none of the
    neighbors of $d_k$ belongs to $I_\forall[V]$. This implies that
    none of the literals using a universal variable is satisfied by
    $V$. Hence there must exist a literal $\ell_{k,p}$ of the clause
    $c_{k,p}$ that uses an existential variables and that is satisfied
    by $V$. Consequently, we have $L_{k,p}\in I_\exists[V]$ and
    $L_{k,p} \succ_\Psi d_k$.

    For the \emph{only if} part take any $k\in\{1,\ldots,s\}$ and
    consider the conjunct $c_k=\ell_{k,1}\lor \ell_{k,2} \lor
    \ell_{k,3}$. If none of the literals, which use universal
    variables, is satisfied by $V$, then none of the corresponding
    $L_{k,p}$ belongs to $I_\forall[V]$, and consequently, $d_k$ is in
    $I_\forall[V]$. Then $I_\exists[V]$ must contain a fact $L_{k,p'}$
    corresponding to one of the literals of $c_k$ using an existential
    variable. This implies that $V\models \ell_{k,p'}$, and
    consequently, $V\models c_k$. \qed
  \end{proof}
  This gives us.
  \begin{corollary}
    The QBF $\Psi$ is valid if and only if for any strict
    $\forall$-repair $I'$ there exists an $\exists$-repair $I''$ such
    that $I''\gg_\mathcal{G} I'$.
  \end{corollary}
  Because only an $\exists$-repair can be preferred over a strict
  $\forall$-repair and for every non-strict $\forall$-repair there is
  a more preferred strict $\forall$-repair, we can make a more general
  statement.
  \begin{corollary}
    The QBF $\Psi$ is valid if and only if for any $\forall$-repair
    $I'$ there exists a repair $I''$ such that $I'' \gg_\mathcal{G}
    I'$.
  \end{corollary}
  $\forall$-repairs are defined as repairs that do not contain the
  fact $p_\exists$ and thus:
  \begin{align*}
    &{}\models \forall x_1,\ldots,x_n.\  \exists
    x_{n+1},\ldots,x_{n+m}.\  \Phi &
    &\text{iff}\\
    &\forall I' \in \Rep(I_\Psi,F).\  [I' \models \neg p_\exists]
    \Rightarrow [\exists I'' \in \Rep(I_\Psi,F).\  I''\gg_\mathcal{G}
    I']&
    &\text{iff}\\
    &\forall I' \in \Rep(I_\Psi,F).\  [\nexists I'' \in
    \Rep(I_\Psi,F).\  I''\gg_\mathcal{G} I'] \Rightarrow [I' \models
    p_\exists]&
    &\text{iff}\\
    &\forall I' \in \GRep(I_\Psi,F,\mathord{\succ}_\Psi) .\ 
    I' \models p_\exists&
    &\text{iff}\\
    &(I_\Psi,\mathord{\succ}_\Psi) \in
    \mathcal{D}_{F,p_\exists}^\mathcal{G}.
  \end{align*}
  We finish by observing that the reduction can be carried out in
  polynomial time.

  To show coNP-hardness of $\mathcal{B}_F^\mathcal{G}$ we remark that
  a 3CNF formula $\Phi$ can be treated as a $\forall^*\exists^*$QBF
  with no universal variables. This way, we use the previous
  transformation to reduce the complement of 3SAT to
  $\mathcal{B}_F^\mathcal{G}$; If $I_\Phi$ is the instance obtained
  in the reduction with $\Phi$, then $\{p_\exists\}$ is a
  globally-optimal repair of $I_\Phi$ if and only if $\Phi\not\in
  \text{3SAT}$.\qed
\end{proof}
\section{Pareto-optimal repairs}
\label{sec:pareto}
The high computational cost of using global optimality compels us
to seek different notions of optimality that may reduce the
computational complexity. The next family of repairs that we consider 
is closely related to $\GRep$. Similarly to $\GRep$, 
it selects a set of repairs whose compliance with the priority 
cannot be further improved by replacing a set of facts with 
a more preferred set of facts. The only difference is in the way 
we lift the priority relation to a preference relation of sets of 
facts. This notion is inspired by the construction of the Pareto 
optimal set of vectors~\cite{KoPa07}.
\begin{definition}[Pareto-optimal repairs $\PRep$]
\label{def:pareto}
Given an instance $I$, a set of integrity constraints $F$, and a
priority $\succ$, an instance $I'\subseteq I$ is \emph{Pareto optimal}
w.r.t.\ $\succ$ and $F$ if no nonempty subset $X$ of facts from $I'$
can be replaced with a nonempty set $Y$ of facts from $I\setminus I'$
such that
\begin{equation}\tag{$\Asterisk_\mathcal{P}$}\label{eq:pareto-optimal}
\forall x \in X.\  \forall y \in Y.\  y \succ x
\end{equation}
and the resulting set of facts is consistent with
$F$. $\PRep$ is the family of Pareto-optimal repairs
i.e., $\PRep(I,F,\mathord{\succ})$ is the set of all repairs
of $I$ w.r.t.\ $F$ that are Pareto optimal w.r.t.\ $\succ$ and $F$.
\end{definition}
We emphasize that the family $\PRep$ selects all
Pareto-optimal repairs. In general, it is, however, possible to define
a family that selects only some of the Pareto-optimal repairs, or even
more generally, a family that constructs a set of Pareto-optimal
instances that need not be repairs. This will allow us to state some
general results e.g., Theorem~\ref{thm:any-pareto-intractable} states
that any family of Pareto-optimal repairs that satisfies
$\mathcal{P}1$ and $\mathcal{P}$ leads inadvertently to intractability
of preferred consistent query answering. In the sequel, we fix an
instance $I$ and a set of denial constraints $F$, and omit them when
referring to the elements of $\PRep(I,F,\mathord\succ)$.
\begin{proposition}\label{prop:pareto-properties}
  $\PRep$ satisfies $\mathcal{P}1$-$\mathcal{P}4$. Also,
  $\GRep \sqsubseteq \PRep$.
\end{proposition}
\begin{proof}
  $\GRep\sqsubseteq\PRep$ follows from
  Definitions~\ref{def:global} and~\ref{def:pareto}. The arguments
  used to prove $\mathcal{P}1$ through $\mathcal{P}4$ are essentially
  the same as in Proposition~\ref{prop:global-properties}. \qed
\end{proof}

\begin{figure}[htb]
\begin{center}\begin{small}
\begin{tikzpicture}[>=stealth']
\node[draw] at (0,0)  (b1) {$\Mgr(Bob,\$70k,RD)$};
\node[draw] at (0,-2) (b2) {$\Mgr(Bob,\$60k,AD)$};

\node[draw] at (4,0)  (m1) {$\Mgr(\Mary,\$50k,PR)$};
\node[draw] at (4,-2) (m2) {$\Mgr(\Mary,\$40k,IT)$};

\node[draw] at (8,0)  (k1) {$\Mgr(\Ken,\$60k,IT)$};
\node[draw] at (8,-2) (k2) {$\Mgr(\Ken,\$50k,PR)$};

\draw[->] (b1) -- (b2);
\draw[->] (k1) -- (k2);
\draw[->] (m1) -- (m2);

\draw[-] (m1) -- (k2);
\draw[-] (k1) -- (m2);
\end{tikzpicture}
\end{small}\end{center}
\caption{\label{fig:preference} Prioritized conflict graph from
  Example~\ref{ex:preference}.}
\end{figure}
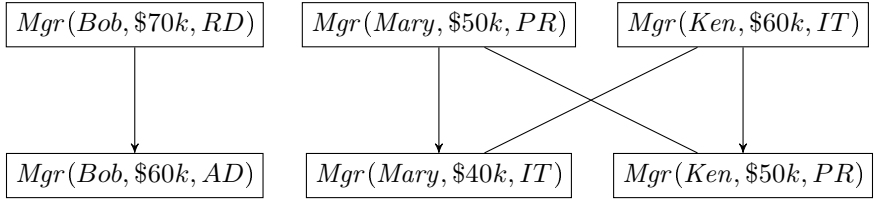

To show that $\PRep\varnot\sqsubseteq\GRep$ we recall
the instance $I_1$ from Example~\ref{ex:preference} whose prioritized
conflict graph is in Figure~\ref{fig:preference}. The repairs $I_1'$
and $I_2'$ are Pareto optimal but only $I_1'$ is globally optimal.  

The family of Pareto-optimal repairs can be viewed as an approximation
of $\GRep$ enjoying better computational properties. We
believe, however, that Pareto optimality is a more cautious and
conservative alternative to global optimality because it requires a
stronger support from the priority to eliminate a repair. For
instance, recall that $I_2'$ from Example~\ref{ex:preference} is not
globally optimal because we can replace $\Mgr(\Mary,\$40k,IT)$ with
the more preferred $\Mgr(\Mary,\$50k,PR)$ and $\Mgr(\Ken,\$50k,PR)$
with the more preferred $\Mgr(\Ken,\$60k,IT)$. However, the same
process can be seen as replacing $\Mgr(\Mary,\$40k,IT)$ with
$\Mgr(\Ken,\$60k,IT)$ and $\Mgr(\Ken,\$50k,PR)$ with
$\Mgr(\Mary,\$50k,PR)$, and neither of those swaps improves the
compliance with the preference. Consequently, $I_2'$ is Pareto
optimal.

Similarly to $\GRep$, $\mathcal{P}$-preferred repairs have an
alternative characterization that is based on extending the priority
to a pre-order on repairs.
\begin{proposition}
  For a given priority $\succ$ and two repairs $I_1'$ and $I_2'$,
  $I_1'$ \emph{Pareto dominates} $I_2'$, denoted
  $I_1'\gg_\mathcal{P}I_2'$ if
\begin{equation}\tag{$\bigvarstar_\mathcal{P}$}\label{eq:pareto-domination}
\exists y\in I_1' \setminus I_2' .\  
\forall x\in I_2' \setminus I_1' .\  
y \succ x.
\end{equation}
The following facts hold:
\begin{itemize}
\item[(i)] a repair $I'$ is Pareto optimal if and only if it is
  $\gg_\mathcal{P}$-maximal i.e., there is no repair $I''$ different
  from $I'$ such that $I''\gg_\mathcal{P} I'$;
\item[(ii)] if $\succ$ is acyclic, then so is $\gg_\mathcal{P}$.
\end{itemize}
\end{proposition}
\begin{proof} \emph{(i)} We prove the contraposition i.e., $I'$ is not
  Pareto optimal if and only if there exists a repair $I''\neq I'$
  such that $I''\gg_\mathcal{P} I'$.

  For the \emph{if} part take $X=I'\setminus I''$ and $Y=\{y\}$, where
  $y\in I''\setminus I'$ such that $\forall x\in X$ we have $y\succ x$
  (it exists by $I''\gg_\mathcal{P} I'$). Clearly, $X$ and $Y$
  validate~\eqref{eq:pareto-optimal} and $(I'\setminus X) \cup Y$ is
  consistent (as a subset of $I''$). Consequently, $I'$ is not Pareto
  optimal. For the \emph{only if} part take any nonempty $X\subseteq
  I'$ and $Y\subseteq I\setminus I'$ such that
  \eqref{eq:pareto-optimal} holds and $J=(I'\setminus X)\cup Y$ is
  consistent. Take any repair $I''$ that contains all facts of
  $J$. Clearly, $I'\setminus I''=X$ and $Y\subseteq I''\setminus I'$
  so it suffices to take $X$ and any $y\in Y$ to verify
  \eqref{eq:pareto-domination}.

  \emph{(ii)} We observe that $I'\gg_\mathcal{P} I''$ implies
  $I'\gg_\mathcal{G} I''$. Thus, if $\gg_\mathcal{P}$ has cycles, then
  so does $\gg_\mathcal{G}$, and consequently, $\succ$. \qed
\end{proof}

The class of Pareto-optimal repairs is the largest class of preferred
repairs we consider in this paper. The remaining families select
subsets of Pareto optimal, and in general, it is possible to consider
other families that select only Pareto-optimal repairs.  The following
result states a rather general observation on the computational
implications of introducing preferences to the framework of consistent
query answers.  
\begin{theorem}\label{thm:any-pareto-intractable}
  There exists an atomic query $Q$ and a set $F$ of two FDs such that
  for any family $\XRep$ of Pareto optimal repairs satisfying
  $\mathcal{P}1$ and $\mathcal{P}2$ the problem of
  $\mathcal{X}$-consistent query answering i.e., the membership of the
  set
  \[
  \mathcal{D}_{F,Q}^{\mathcal{X}}=\{
  (I,\mathord{\succ}) \sep 
  \forall I'\in\XRep(I,F,\mathord{\succ}).\  I'\models Q 
  \},
  \]
  is coNP-hard.
\end{theorem}
\begin{proof}
  We show the hardness by reducing the complement of SAT to
  $\mathcal{D}_{F,Q}^\mathcal{X}$. Take then any CNF formula
  $\Phi=c_1\land \ldots{} \land c_k$ over variables $x_1,\ldots{},x_n$
  and let $c_j=\ell_{j,1}\lor \ldots \lor \ell_{j,m_j}$. We assume
  that there are no repetitions of literals in a clause (i.e.,
  $\ell_{j,k_1} \neq \ell_{j,k_2}$). We construct a relation instance
  $I_\Phi$ over the schema $R(A_1,B_1,A_2,B_2)$ in the presence of two
  functional dependencies $F=\{A_1\rightarrow B_1, A_2\rightarrow
  B_2\}$. The instance $I_\Phi$ consists of the following facts:
  \begin{itemize}
  \item $w_i=R(i,1,i,1)$ corresponding to the positive valuation of
    $x_i$ (for $i\in\{1,\ldots{},n\}$),
  \item $\bar{w}_i=R(i,-1,-i,1)$ corresponding to the negative
    valuation of $x_i$ (for every $i\in\{1,\ldots{},n\}$),
  \item $d_j=R(n+j,1,0,1)$ corresponding to the clause $c_j$ (for
    every $j\in\{1,\ldots,m\}$),
  \item $v_i^j=R(n+j,1,-i,0)$ encoding the use of $x_i$ in the clause
    $c_j$ (for any $i\in\{1,\ldots,n\}$ and $j\in\{1,\ldots,m\}$ such
    that $c_j$ uses $x_i$),
  \item $\bar{v}_i^j=R(n+j,1,i,0)$ encoding the use of $\neg x_i$ in
    the clause $c_j$ (for any $i\in\{1,\ldots,n\}$ and
    $j\in\{1,\ldots,m\}$ such that $c_j$ uses $\neg x_i$),
  \item $b=R(0,0,0,0)$ corresponding to the formula $\Phi$.
  \end{itemize}
  The constructed priority $\succ_\Phi$ is the minimal priority of
  $I_\Phi$ w.r.t.\ $F$ such that:
  \begin{align*}
    & \bar{w}_i \succ_\Phi v_i^j,& & v_i^j \succ_\Phi d_j, &
    & d_j\succ_\Phi b,\\
    & w_i \succ_\Phi \bar{v}_i^j,& & \bar{v}_i^j \succ_\Phi d_j.
  \end{align*}
  Figure~\ref{fig:pareto-reduction} presents prioritized conflict
  graph obtained from the formula $\Phi = (\neg x_1 \lor x_2 \lor x_3)
  \land (\neg x_3 \lor \neg x_4 \lor x_5).$
  \begin{figure}[htb]
    \begin{center}
      \begin{tikzpicture}[yscale=1.15,xscale=0.95]
        \path[use as bounding box] (.75,.5) rectangle (12.25,3.75);
        \path (0,3.5) ++(1,0) node (nw1) {$\bar{w}_1$} ++(1,0) node
        (w1) {$w_1$} ++(1,0) node (nw2) {$\bar{w}_2$} ++(1,0) node
        (w2) {$w_2$} ++(1,0) node (nw3) {$\bar{w}_3$} ++(1,0) node
        (w3) {$w_3$} ++(1,0) node (nw4) {$\bar{w}_4$} ++(1,0) node
        (w4) {$w_4$} ++(1,0) node (nw5) {$\bar{w}_5$} ++(1,0) node
        (w5) {$w_5$} ++(1,0) node (nw6) {$\bar{w}_6$} ++(1,0) node
        (w6) {$w_6$}; \path (2,2.5) node (nv11) {$\bar{v}_1^1$}; \path
        (3,2.5) node (v21) {$v_2^1$}; \path (4.5,2.5) node (v31)
        {$v_3^1$}; \path (5.5,2.5) node (v32) {$v_3^2$}; \path (8,2.5)
        node (nv42) {$\bar{v}_4^2$}; \path (9,2.5) node (v52)
        {$v_5^2$}; \path (12,2.5) node (nv62) {$\bar{v}_5^2$}; \path
        (0,1.5) ++(4,0) node (d1) {$d_1$} ++(4.5,0) node (d2) {$d_2$};
        \path (0,0.75) ++(6.5,0) node (b) {$b$}; \draw (nw1) -- (w1)
        (nw2) -- (w2) (nw3) -- (w3) (nw4) -- (w4) (nw5) -- (w5) (nw6)
        -- (w6); \draw[-latex] (w1) -- (nv11); \draw[-latex] (nw2) --
        (v21); \draw[-latex] (nw3) -- (v31); \draw[-latex] (nw3) --
        (v32); \draw[-latex] (w4) -- (nv42); \draw[-latex] (nw5) --
        (v52); \draw[-latex] (w6) -- (nv62); \draw [-latex] (nv11) --
        (d1); \draw [-latex] (v21) -- (d1); \draw [-latex] (v31) --
        (d1); \draw [-latex] (v32) -- (d2); \draw [-latex] (nv42) --
        (d2); \draw [-latex] (v52) -- (d2); \draw [-latex] (nv62) --
        (d2); \draw[latex-] (b) -- (d1); \draw[latex-] (b) -- (d2);
      \end{tikzpicture}
    \end{center}
    \caption{\label{fig:pareto-reduction} The prioritized conflict
      graph for $\Phi= (\neg x_1 \lor x_2 \lor x_3) \land (\neg x_3
      \lor \neg x_4 \lor x_5).$}
  \end{figure}
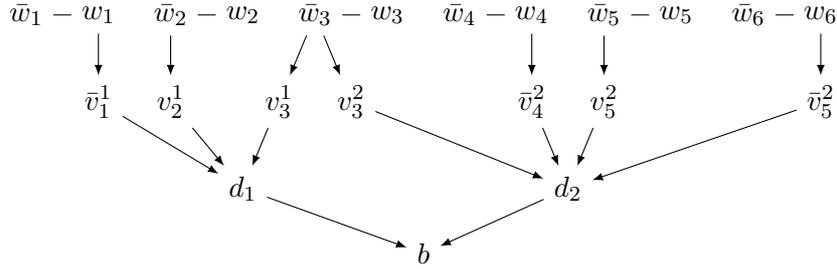
  The query we consider is $Q=\neg b$. We claim that
  \[
  (I_\Phi,\mathord\succ_\Phi) \in \mathcal{D}_{F,Q}^\mathcal{X} \iff
  \forall I' \in \XRep(I_\Phi,F,\mathord\succ_\Phi).\  b \not
  \in I' \iff \Phi \not \in \text{SAT}.
  \]
  For the \emph{if} part, suppose there exists a repair
  $I'\in\XRep(I_\Phi,F,\mathord\succ_\Phi)$ such that $b\in
  I'$. Obviously, for every $j\in\{1,\ldots,m\}$ the fact $d_j$ does
  not belong to $I'$. Also, for every $j$ at least one fact
  neighboring to $d_j$, other than $b$, is present in $I'$, or
  otherwise $I'$ is not a Pareto-optimal repair. Similarly, $I'$ has
  either $w_i$ or $\bar{w}_i$ for every $i\in\{1,\ldots,n\}$, and
  hence, the following valuation is properly defined:
  \[
  V(x_i) =
  \begin{cases}
    \true & \text{if $w_i\in I'$,}\\
    \false & \text{if $\bar{w}_i \in I'$.}
  \end{cases}
  \]
  We claim that $V\models\Phi$. Suppose otherwise and take any clause
  $c_j$ unsatisfied by $V$. Let $x\neq b$ be the fact neighboring to
  $d_j$ that is present in $I'$. W.l.o.g.\ we can assume that $x =
  \bar{v}_{j,i_0}$ for some $i_0$ and then $\neg x_{i_0}$ is a literal
  of $c_j$. Also then, $w_{i_0}$ does not belong to $I'$ and so
  $V(x_{i_0})=\false$. This implies that $V\models\neg x_{i_0}$ and
  $V\models c_j$; a contradiction.

  For the \emph{only if} part, suppose there exists a valuation $V$
  such that $V\models \Phi$ and consider the following instance
  \begin{align*}
    I'=\{&w_i \sep V(x_i)=\true\} \cup
    \{\bar{w}_i \sep V(x_i)=\false\} \cup{}\\
    \{&v_i^j \sep V(x_i)=\true\} \cup \{\bar{v}_i^j \sep
    V(x_i)=\false\} \cup \{b\}.
  \end{align*}
  First, we note that $I'$ is a repair and a Pareto-optimal one. Next,
  we show that $I'\in\XRep(I_\Phi,F,\mathord\succ_\Phi)$. To
  prove this consider the following priority
  $\mathord\succ'=\mathord\succ_\Phi\cup \{(v_i,\bar{v}_i) \sep
  V(x_i)=\true\} \cup\{(\bar{v}_i,v_i) \sep V(x_i)=\false\}$.  It can
  be easily verified that $I'$ is the only Pareto-optimal repair of
  $I_\Phi$ w.r.t.\ $F$ and $\succ'$. Since $\XRep$ satisfies
  $\mathcal{P}1$, we get
  $I'\in\XRep(I_\Phi,F,\mathord\succ')$. Note that $\succ'$
  is an extension of $\succ_\Phi$ and thus $I'$ belongs to
  $\XRep(I_\Phi,F,\mathord\succ)$ by $\mathcal{P}2$. Finally,
  we observe that $b\in I'$ which implies that $\true$ is not an
  $\mathcal{X}$-preferred consistent query answer to $Q$ in $I_\Phi$
  w.r.t.\ $F$ and $\succ_\Phi$; a contradiction.

  We finish the proof with the observation that the described
  reduction requires time polynomial in the size of the formula
  $\Phi$. \qed
\end{proof}
We also present an alternative characterization of Pareto-optimal
repairs that yields a tractable procedure for repair checking.
\begin{lemma}\label{lemma:pareto-check}
  A repair $I'$ is not Pareto optimal w.r.t.\ $\succ$ if and only if
  there exists a fact $y\in I\setminus I'$ such that for every
  conflict $C$ in $I'\cup\{y\}$ there is $x\in C$ such that $y\succ
  x$.
\end{lemma}
\begin{proof}
  For the \emph{if} part, let $C_1,\ldots,C_k$ be all conflicts in
  $I'\cup\{y\}$ and $x_i$ be the element of $C_i$ such that $y\succ
  x_i$ (for $i\in\{1,\ldots,k\}$). Clearly, $(I'\setminus
  \{x_1,\ldots,x_k\})\cup \{y\}$ is consistent, which shows that $I'$
  is not Pareto optimal.

  For the \emph{only if} part, take any nonempty $X$ and $Y$ such that
  $(I'\setminus X)\cup Y$ is consistent and $\forall y\in Y.\ \forall
  x\in X.\  y\succ x$. Fix any $y\in Y$ and take any conflict $C$ in
  $I'\cup\{y\}$. Clearly, $C$ contains an element $x$ of $X$ since
  $(I'\setminus X)\cup Y$ is consistent. Naturally $y\succ x$. \qed
\end{proof}

\begin{corollary}\label{cor:pareto-intractable}
$\mathcal{P}$-preferred repair checking is in LOGSPACE and 
$\mathcal{P}$-preferred consistent query answering is coNP-complete.  
\end{corollary}
\begin{proof}
  We observe that to check the condition of
  Lemma~\ref{lemma:pareto-check} we need to iterate over $I\setminus
  I'$ which can be accomplished with two pointers: one to iterate over
  $I$ and the other to scan $I'$. Recall that a conflict is a set of
  facts and its cardinality is bounded by the size of $F$ which is
  assumed to be a constant parameter. Hence, we can iterate over all
  conflicts of $I'$ (extended with one fact) using a constant number
  of pointers scanning $I'$. Consequently, $\mathcal{P}$-preferred
  repair checking is in LOGSPACE. $\mathcal{D}_{F,Q}^\mathcal{P}$
  belongs to coNP from the definition of $\mathcal{P}$-preferred
  consistent query answers and is coNP-complete by
  Theorem~\ref{thm:any-pareto-intractable}. \qed
\end{proof}

Now, we investigate a sound and complete algorithm for computing
Pareto-optimal repairs. First, we observe that it is possible to use
an algorithm similar to Algorithm~\ref{alg:global-repairing}, starting
with an arbitrary repair and attempting to iteratively improve its
compliance with the priority until a Pareto-optimal repair is
reached. While checking Pareto optimality can be done in polynomial
time, we note that the sequence of repairs, constructed in
Example~\ref{ex:exp-chain}, of exponential length is also a
$\gg_\mathcal{P}$-chain. Consequently, such an algorithm may require
an exponential number of iterations to obtain a Pareto-optimal repair.

We propose a simpler approach where we construct an arbitrary repair
and if it is not Pareto optimal we discard it and construct a
completion-optimal repair using Algorithm~\ref{alg:prioritized-repairing}
presented in the next section.  Completion-optimal repairs constitute a
subset of Pareto-optimal repairs and thus if the algorithm fails to
construct a Pareto-optimal repair in the first stage, then the repair
constructed in the second stage is Pareto optimal. Consequently,
Algorithm~\ref{alg:pareto-repairing} is sound. We recall that
Algorithm~\ref{alg:repairing} is complete, i.e it may return any
repair, in particular, any Pareto-optimal repair. If the repair
constructed in step~1 of Algorithm~\ref{alg:pareto-repairing} is
Pareto optimal, then this repair is returned. Consequently, the
algorithm is complete. Finally, it works in polynomial time since
checking Pareto optimality is in LOGSPACE and
Algorithms~\ref{alg:repairing} and~\ref{alg:prioritized-repairing}
work in polynomial time.
\begin{algorithm}[htb]
\caption{\label{alg:pareto-repairing} Constructing a Pareto-optimal
    repair of $I$ w.r.t.\ $F$ and $\succ$.}
\begin{tabbing}
mm\=mm\=mm\=mm\=\kill
\step{1} \> {\bf construct a repair} $I'$ of $I$\quad \emph{/*Algorithm~\ref{alg:repairing}*/}\\
\step{2} \> {\bf if} $I'$ is Pareto optimal w.r.t.\ $\succ$ {\bf then}\\
\step{3} \> \> {\bf return} $I'$ \\
\step{4} \> {\bf else } \\
\step{5} \> \> {\bf return} any completion-optimal repair of $I$ w.r.t.\ $\succ$ \quad \emph{/*Algorithm~\ref{alg:prioritized-repairing}*/}
\end{tabbing}%
\vspace{-9pt}
\end{algorithm}
\begin{proposition}
  Algorithm~\ref{alg:pareto-repairing} is a sound and complete
  algorithm constructing Pareto-optimal repairs. It works in time
  polynomial in the size of the input instance and the priority
  relation.
\end{proposition}
\section{Completion-optimal repairs}
\label{sec:common}
The last family of preferred repairs is based on a notion of
optimality different from global and Pareto optimality and intuitively
can be described as follows. When repairing a database with a priority
that is not total and resolving a conflict that is not prioritized, we
commit to a particular prioritization of this conflict. In this view,
constructing a repair that conforms to a given priority is equivalent
to constructing a total extension of that priority such that the
constructed repair is the only repair globally optimal w.r.t.\ the
total priority. We remark that this notion is quite robust as it
remains identical if we replace in it global optimality by Pareto
optimality. The same holds for all results stated in this
section. This is because $\GRep$ and $\PRep$
coincide for total priorities by
$\GRep\sqsubseteq\PRep$ and $\mathcal{P}4$ for
$\PRep$ and $\GRep$. Another motivation for the
family of repairs presented in this section is a fairly intuitive and
natural repairing algorithm which we present later on.
\begin{definition}[Completion-optimal repairs $\CRep$]
  Given an instance $I$, a set of denial constraints $F$, and a
  priority $\succ$, an instance $I'\subseteq I$ is \emph{completion
    optimal} w.r.t.\ $\succ$ and $F$ if and only if there exists a
  total priority $\mathord{\succ}'\supseteq\mathord{\succ}$ such that
  $I'$ is globally optimal w.r.t.\ $\succ'$ and $F$. $\CRep$
  is the family of completion-optimal repairs
  i.e., $\CRep(I,F,\mathord\succ)$ is the set of all repairs
  of $I$ w.r.t.\ $F$ that are completion optimal w.r.t.\ $\succ$ and $F$.
\end{definition}
We remark that $\CRep$ selects all completion-optimal
repairs and that it is possible to consider families that select only
some completion-optimal repairs. We fix an instance $I$ and a set of
denial constraints $F$, and omit them when referring to the elements
of $\CRep(I,F,\mathcal\succ)$.
\begin{example}\label{ex:comm-optim-repa}
Consider the schema of one relation name $R(A,B,C,D)$ with a set of
two functional dependencies $F_4=\{R:A\rightarrow B, R:C\rightarrow
D\}$. Take the following instance
\[
I_4=\{R(1,1,1,1),R(1,2,1,2),R(1,3,0,0),R(0,0,1,3)\}
\]
and the following priority relation
\[
\mathord{\succ}_4=\{(R(1,1,1,1),R(1,3,0,0)), (R(1,2,1,2),R(0,0,1,3))\}.
\] 
The corresponding prioritized conflict graph is presented in
Figure~\ref{fig:crep-not-grep}.
\begin{figure}[htb]
\begin{center}\begin{small}
\begin{tikzpicture}[>=stealth']
\node[draw] at (0,0) (t1) {$R(1,1,1,1)$};
\node[draw] at (3,0) (t2) {$R(1,2,1,2)$};
\node[draw] at (0,-2) (t3) {$R(1,3,0,0)$};
\node[draw] at (3,-2) (t4) {$R(0,0,1,3)$};

\draw[-] (t1) -- (t2);
\draw[-] (t2) -- (t3);
\draw[-] (t1) -- (t4);
\draw[->] (t1) -- (t3);
\draw[->] (t2) -- (t4);
\end{tikzpicture}
\end{small}\end{center}
\caption{\label{fig:crep-not-grep}The prioritized conflict graph
  $G(I_4,F_4,\mathord{\succ}_4)$.} 
\end{figure}
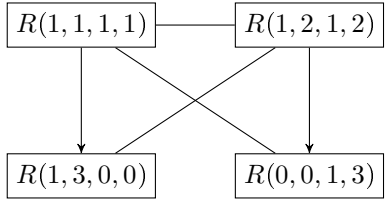
The instance $I_4$ has 3 repairs: 
\begin{align*}
& I_1' =\{R(1,1,1,1)\}, &
& I_2' =\{R(1,2,1,2)\}, &
& I_3' =\{R(1,3,0,0), R(0,0,1,3)\}. 
\end{align*}
We note that all three repairs are globally optimal w.r.t.\ $\succ_4$. 
The repairs $I_1'$ and $I_2'$ are completion optimal as witnessed by the
following total extensions of $\succ_4$ ($\succ_4'$ for $I_1'$ and
$\succ_4''$ for $I_2'$):
\begin{align*}
\mathord\succ_4'= \mathord\succ_4\cup\{
(R(1,1,1,1),R(0,0,1,3)),
(R(1,2,1,2),& R(1,3,0,0)),\\
&(R(1,1,1,1),R(1,2,1,2))
\}\\
\mathord\succ_4''= \mathord\succ_4\cup\{
(R(1,1,1,1),R(0,0,1,3)),
(R(1,2,1,2),& R(1,3,0,0)),\\
&(R(1,2,1,2),R(1,1,1,1))
\}.
\end{align*}
On the other hand, there is no total extension of $\succ_4$ for which
the repair $I_3'$ is globally optimal, and hence, $I_3'$ is not
completion optimal.\qed
\end{example}

It is an open question whether there exists an intuitive definition of
a pre-order on repairs whose maximal elements are exactly
completion-optimal repairs. We show, however, the family of
completion-optimal repairs is the smallest family of globally-optimal
repairs that satisfies the properties $\mathcal{P}1$ and
$\mathcal{P}2$.  Notice that $\GRep$ is only one of the
possible families of globally-optimal repairs satisfying
$\mathcal{P}1$ and $\mathcal{P}2$.  $\CRep$ is another one.
\begin{lemma}\label{lemma:common-acyclic-extensions}
  $\CRep\sqsubseteq \XRep$ for every family
  $\XRep$ of globally-optimal repairs that satisfies
  $\mathcal{P}1$ and $\mathcal{P}2$. In other words, a repair $I'$ is
  completion optimal w.r.t.\ $F$ and $\succ$ if and only if
  $I'\in\XRep(I,F,\mathord\succ)$ for every family
  $\XRep$ of globally-optimal repairs that satisfies
  $\mathcal{P}1$ and $\mathcal{P}2$. 
\end{lemma}
\begin{proof}
  For the \emph{only if} part, observe that by $\mathcal{P}4$ for
  $\GRep$ $I'$ is the only globally-optimal repair
  w.r.t.\ $\succ$. Consequently,
  $I'\in\XRep(I,F,\mathord{\succ}')$ for any family
  $\XRep$ satisfying $\mathcal{P}1$. Moreover,
  $I'\in\XRep(I,F,\mathord{\succ})$ because $\XRep$
  satisfies $\mathcal{P}2$ and
  $\mathord{\succ}\subseteq\mathord{\succ}'$. Thus, $I'$ is a
  completion-optimal repair of $I$ w.r.t.\ $F$ and $\succ$.

  For the \emph{if} part, suppose $\succ$ has no acyclic total
  extension $\succ'$ for which $I'$ is globally optimal w.r.t.\
  $\succ'$. Consider the following family of globally-optimal repairs
  \[
  \XRep(I^o,F^o,\mathord{\succ}^o) =
  \begin{cases}
    \GRep(I^o,F^o,\mathord{\succ}^o) \setminus \{I'\} &
    \text{if $\mathord{\succ}^o\supseteq\mathord{\succ}$, $I^o=I$, and $F^o=F$,}\\
    \GRep(I^o,F^o,\mathord{\succ}^o) & \text{otherwise.}
  \end{cases}
  \]
  It can be easily seen that $\XRep$ satisfies $\mathcal{P}1$
  and $\mathcal{P}2$. We observe that
  $I'\not\in\XRep(I,F,\mathord\succ)$. Consequently, $I'$ is
  not a completion-optimal repair of $I$ w.r.t.\ $F$ and $\succ$. \qed
\end{proof}
\begin{proposition}\label{prop:common-properties}
  $\CRep$ satisfies
  $\mathcal{P}1$-$\mathcal{P}4$ and 
  $\CRep\sqsubseteq\GRep$.
\end{proposition}
\begin{proof}
  $\CRep\sqsubseteq\GRep$ because $\GRep$
  is a family of globally-optimal repairs that satisfies both
  $\mathcal{P}1$ and $\mathcal{P}2$
  (cf.\ Proposition~\ref{prop:global-properties}).

  $\mathcal{P}1$ follows from the definition of completion-optimal
  repairs and the observation that any priority $\succ$ can be
  extended to some total $\succ'$ (the same argument as in the proof
  of Proposition~\ref{prop:global-properties}). Therefore,
  $\varnothing\neq\GRep(I,F,\mathord\succ')\subseteq\CRep(I,F,\mathord\succ)$
  by $\mathcal{P}4$ for $\GRep$. $\mathcal{P}2$ follows
  directly from Lemma~\ref{lemma:common-acyclic-extensions}.

  To show $\mathcal{P}3$ we take an arbitrary repair $I'$ and
  construct a priority $\succ$ such that $I'$ is globally optimal
  w.r.t.\ $\succ$. For that we take any total ordering $\succ_1$ of
  $I'$ and any total ordering of $\succ_2$ of $I\setminus I'$. We
  obtain $\succ$ by a diligent composition of $\succ_1$ with
  $\succ_2$:
  \[
  R(t) \succ R'(t') \iff
  \begin{cases}
    R(t) \succ_1 R'(t') & \text{if $R(t),R'(t')\in I'$,}\\
    \true & \text{if $R(t)\in I'$ and $R'(t')\in I\setminus I'$,}\\
    R(t) \succ_2 R'(t') & \text{if $R(t),R'(t')\in I\setminus I'$,}\\
    \false & \text{if $R(t)\in I\setminus I'$ and $R'(t')\in I'$,}
  \end{cases}
  \]
  for any two neighboring facts $R(t)$ and $R'(t')$ ($R(t)\not\succ
  R'(t')$ if $R(t)$ and $R'(t')$ are not neighboring). Clearly,
  $\succ$ is acyclic since it is based on the acyclic components
  $\succ_1$ and $\succ_2$, and we add an element $(R(t),R'(t'))$ only
  if $R(t)\in I'$ and $R'(t')\not\in I'$. Naturally, $\succ$ is a
  total priority. It is also easy to verify that $I'$ is globally
  optimal w.r.t.\ $\succ$. $\mathcal{P}4$ follows from
  $\CRep\sqsubseteq\GRep$, $\mathcal{P}4$ for
  $\GRep$, and $\mathcal{P}1$ for $\CRep$ proved
  above.  \qed
\end{proof}

Completion-optimal repairs can be also characterized as exactly those
repairs that can be obtained with an iterative accumulation of facts
selected with the \emph{winnow operator} \cite{ChTODS03}:
\[
\omega_{\mathord\succ}(I)=\{R(t)\in I \sep\nexists R'(t')\in I.\  R'(t')\succ R(t)\}.
\]
\begin{algorithm}[htb]
\caption{\label{alg:prioritized-repairing} Constructing a completion-optimal repair.}
\begin{tabbing}
mm\=mm\=mm\=mm\=\kill
\step{1} \> $I^o \gets I$ \\
\step{2} \> $J\gets\varnothing$ \\
\step{3} \> {\bf while} $\omega_{\mathord\succ}(I^o)\neq\varnothing$ {\bf do}\\
\step{4} \> \> {\bf choose} $R(t)\in \omega_{\mathord\succ}(I^o)$\\
\step{5} \> \> $I^o\gets I^o\setminus\{R(t)\}$\\
\step{6} \> \> {\bf if } $J\cup\{R(t)\}\models F$ {\bf then}\\
\step{7} \> \> \> $J\gets J\cup\{R(t)\}$ \\
\step{8} \> {\bf return} $J$
\end{tabbing}%
\vspace{-9pt}
\end{algorithm}
\begin{theorem}
  Algorithm~\ref{alg:prioritized-repairing} is a sound and complete
  algorithm constructing completion-optimal repairs. It works in time
  polynomial in the size of the input instance and the priority
  relation.
\end{theorem} 
\begin{proof}
  We observe that the instance resulting from an execution of
  Algorithm~\ref{alg:prioritized-repairing} can be associated with the
  \emph{sequence of choices} made in line~4 during the execution. We
  also observe that this sequence is an ordering of the facts of the
  original instance $I$.

  To show soundness, we take an instance $I'$ obtained with the
  sequence of choices $x_1,\ldots,x_n$. We show that $I'$ is completion
  optimal by extending $\succ$ to a total priority $\succ'$ for which
  $I'$ is globally optimal. The priority $\succ'$ is defined as
  \[
  x_i \succ' x_j \iff \text{$x_i$ and $x_j$ are neighboring and
    $i<j$}.
  \]
  Clearly, $\succ'$ is acyclic and a total priority. We also observe
  that $\mathord{\succ}\subseteq\mathord{\succ}'$ because $x_i \succ
  x_j$ implies that $i<j$ i.e., $x_i$ is selected before $x_j$ and the
  choices are constrained by $\omega_{\mathord{\succ}}$.

  To show that $I'$ is globally optimal w.r.t.\ $\succ'$ take any
  $X\subseteq I'$ and any $Y\subseteq I\setminus I'$ such that
  $(I'\setminus X)\cup Y$ is consistent. Now, take any $x_j\in Y$ and
  observe that adding $x_j$ to the instance being created by
  Algorithm~\ref{alg:prioritized-repairing} must have been prevented
  by some conflict $\{x_{i_1},\ldots,x_{i_k},x_j\}$ with the facts
  added previously i.e., $i_\ell < j$ for
  $\ell\in\{1,\ldots,k\}$. Consequently, $x_{i_\ell}\succ' x_j$ for
  $\ell\in\{1,\ldots,k\}$. We observe that at least one of
  $x_{i_1},\ldots,x_{i_k}$ must be present in $X$ since $Y$ contains
  $x_j$, $I'$ contains $x_{i_1},\ldots,x_{i_k}$, and $(I'\setminus
  X)\cup Y$ is consistent. Thus, $x_j \not\succ x_\ell$ for some
  $x_\ell\in X$, $I'$ is globally optimal w.r.t.\ $\succ'$, and by
  $\mathcal{P}2$ for $\GRep$ we get that $I'$ is globally
  optimal w.r.t.\ $\succ$.

  To show completeness, we take a completion-optimal repair $I'$ and the
  total priority $\succ'$ for which $I'$ is globally optimal and use
  $\succ'$ to construct a valid sequence of choices yielding
  $I'$. Naturally, the same choice sequence is valid for an execution
  with $\succ$ because $\succ'$ extends $\succ$.

  Take an execution of Algorithm~\ref{alg:prioritized-repairing} on
  $I$ with $\succ'$ that constructs some instance $I''$ with the
  sequence of choices $x_1,\ldots,x_n$. Note that if $x_i$ and $x_j$
  are neighboring, then $x_i \succ' x_j$ if and only if $i <
  j$. Suppose that $I''\neq I'$ and take the minimal index $i$ of the
  element $x_i$ on which $I'$ and $I''$ differ. Note that
  $I'\cap\{x_1,\ldots,x_{i-1}\}=I''\cap\{x_1,\ldots,x_{i-1}\}$ and
  either $x_i\in I'$ and $x_i\not\in I''$, or $x_i\not\in I'$ and
  $x_i\in I''$. The first case is not possible because
  Algorithm~\ref{alg:prioritized-repairing} would have discarded $x_i$
  only if there had been a conflict involving $x_i$ and some facts of
  $I''\cap\{x_1,\ldots,x_{i-1}\}$. Then, however, the same conflict
  would have been included in $I'$ i.e., $I'$ would have not been
  consistent. Suppose then, $x_i\not\in I'$ and $x_i\in I''$. Let
  $C_1,\ldots,C_k$ be all conflicts present in $I'\cup\{x_i\}$ w.r.t.\
  $F$, and since $I'\cup\{x_i\}$ is not consistent, there exists at
  least one conflict in $I'\cup\{x_i\}$. Naturally,
  $I'\cap\{x_1,\ldots,x_{i-1}\}\cup\{x_i\}$ is consistent, and thus
  for every $j\in\{1,\ldots,k\}$ the conflict $C_j$ contains a fact
  $x_{i_j}$ such that $i_j > i$. Let $X=\{x_{i_1},\ldots,x_{i_k}\}$
  and $Y=\{x_i\}$, and observe that $(I'\setminus X) \cup Y$ is
  consistent. Moreover, $X$ and $Y$ satisfy
  \eqref{eq:global-optimality} (Definition~\ref{def:global}) since
  $i_j > i$ implies that $x_i\succ x_{i_j}$. Consequently, $I'$ is not
  globally optimal; a contradiction. \qed
\end{proof}

\begin{corollary}
$\mathcal{C}$-preferred repair checking is in PTIME and 
$\mathcal{C}$-preferred consistent query answering is coNP-complete.  
\end{corollary}
\begin{proof}
  To check if a repair $I'$ is completion optimal we use
  Algorithm~\ref{alg:prioritized-repairing} to simulate the
  construction of $I'$ by restricting the choice in line~4 to facts
  $\omega_{\mathord{\succ}}(J)\cap I'$. It can be easily shown that
  the repair $I'$ is completion optimal if and only if such a simulation
  can be performed successfully (i.e., it produces $I'$).  Naturally,
  $\mathcal{D}_{F,Q}^\mathcal{C}$ belongs to coNP and its
  coNP-completeness follows from
  Theorem~\ref{thm:any-pareto-intractable}. \qed
\end{proof}
The exact complexity of $\mathcal{C}$-preferred repair checking,
namely whether it is PTIME-complete or in LOGSPACE, remains an open
question.

The introduced families of preferred repairs create a hierarchy: 
\[
\CRep\sqsubseteq\GRep\sqsubseteq\PRep.
\]
Recall from the previous section that
$\PRep\neq\GRep$
(cf.\ Figure~\ref{fig:preference}). We note that in
Example~\ref{ex:comm-optim-repa} all repairs are globally optimal but
only $I_1'$ and $I_2'$ are completion optimal which shows that
$\CRep\neq \GRep$. Thus, the hierarchy is proper.
We observe, however, that under certain conditions this hierarchy
collapses.
\begin{proposition}\label{prop:collapse}
  $\PRep$, $\GRep$, and $\CRep$ coincide
  under one of the following conditions:
\begin{itemize}
\item[(i)] the set of constraints $F$ consists of one key dependency only;
\item[(ii)] the priority $\succ$ can be extended to acyclic priorities only. 
\end{itemize}
Moreover, $\GRep$ and $\CRep$ coincide if 
\begin{itemize}
\item[(iii)] the set of constraints $F$ consists of one functional dependency only.
\end{itemize}
\end{proposition}
\begin{proof}
  For \emph{(i)} to show that
  $\PRep\sqsubseteq\CRep$ in the presence of exactly
  one key dependency, we use the fact that the conflict graph is a
  union of pairwise disjoint cliques and every repair consists of
  exactly one element selected from each clique.

  We fix an instance $I$, a key dependency $F$, and a priority
  $\succ$. Let $C_1,\ldots,C_n$ be the cliques of $G(I,F)$. Take any
  $I'\in\PRep(I,F,\mathord\succ)$ and let
  $R_1(t_1),\ldots,R_n(t_n)$ be the elements of $I'$ such that
  $R_i(t_i)\in C_i$. Since $I'$ is Pareto optimal, then for every $i$
  there is no $y\in C_i\setminus\{R(t_i)\}$ such that $y\succ
  R_i(t_i)$, and consequently,
  $R_i(t_i)\in\omega_{\mathord\succ}(C_i)$. Hence,
  $R_1(t_1),\ldots,R_n(t_n)$ is a proper choice sequence for
  Algorithm~\ref{alg:prioritized-repairing}. Finally, we observe that
  if the fact $R_i(t_i)$ has been added to the constructed repair,
  then none of the facts of $C_i\setminus\{R_i(t_i)\}$ can be further
  added.

  For {\it (ii)} We take any $I'\in\PRep(I,F,\mathord\succ)$
  and construct a total extension $\succ'$ of $\succ$ by prioritizing
  in favor of $I'$ all conflicts unprioritized by $\succ$ i.e.,
  $\succ'$ is any total priority such that for any $x\in I'$ and any
  $y$ conflicting with $x$ if $y \not\succ x$ then $x \succ' y$. Since
  $\succ$ can be extended to acyclic orientations only, $\succ'$ is
  acyclic. Clearly, $I'$ is a Pareto-optimal repair w.r.t.\ $\succ'$
  and a unique one by $\mathcal{P}4$ for $\PRep$. Therefore
  $I'\in\CRep(I,F,\mathord\succ')$ and by $\mathcal{P}2$ we
  get $I'\in\CRep(I,F,\mathord\succ)$.

  For {\it (iii)} we assume a single relation name $R$ with the
  functional dependency $X\rightarrow Y$ and use the notions of
  $X$-cluster and $(X,Y)$-cluster (Section~\ref{sec:clusters},
  page~\pageref{sec:clusters}). Let the instance $I$ be the union of
  the $X$-clusters $C_1,\ldots,C_n$. Take any globally-optimal repair
  $I'$ and let it be the union of the $(X,Y)$-clusters
  $D_1,\ldots,D_n$ ($D_i\subseteq C_i$ for every
  $i\in\{1,\ldots,n\}$). By global optimality of $I'$ we have that for
  every $i\in\{1,\ldots,n\}$
  \[
  \exists R_i(t_i)\in D_i.\forall y\in C_i\setminus D_i. y \not \succ
  R_i(t_i).
  \]
  Therefore, Algorithm~\ref{alg:prioritized-repairing} can perform the
  first $n$ iterations with a choice sequence beginning with
  $R_i(t_1),\ldots,R_n(t_n)$. Because $n(R_i(t_i))=C_i\setminus D_i$
  and elements of $D_i$ conflict only with elements of $C_i\setminus
  D_i$, the remaining choices can consist of any ordering of
  $(D_1\setminus\{R_i(t_1)\})\cup\ldots\cup(D_n\setminus\{R_i(t_n)\})$.
  Consequently, $I'$ is a result of
  Algorithm~\ref{alg:prioritized-repairing}.\qed
\end{proof}
We note that the conditions are sufficient but not necessary e.g., the
hierarchy trivially collapses for any set of denial constraints and an
empty priority relation.
\section{Tractable case}
\label{sec:complexity}
The intractability proofs for consistent query answering use at least
2 FDs. Next, we investigate the case when only one FD is present. We
begin by considering queries that are conjunctions of ground literals,
and next, we generalize this approach to arbitrary ground queries.

We observe that if only functional dependencies are considered, facts
can create conflicts only with facts of the same relation, and
therefore, we can limit our consideration to a schema consisting of
one relation name only. Consequently, we assume a single relation name
$R$ with the FD $R:X\rightarrow Y$ and use the notions of $X$-cluster and
$(X,Y)$-cluster (Section~\ref{sec:clusters},
page~\pageref{sec:clusters}). We fix an instance $I$ and a priority
$\succ$. For every fact $R(t)\in I$, by $C_{R(t)}$ we denote the
$X$-cluster to which the fact $R(t)$ belongs to and by $D_{R(t)}$ we
denote its $(X,Y)$-cluster. We also fix the query 
\[
\Phi=R(t_1) \land \ldots{} \land R(t_k) \land \neg R(t_{k+1}) \land
\ldots{} \land \neg R(t_m).
\]
We assume that the facts $R(t_1),\ldots,R(t_k)$ belong to $I$;
otherwise there is no repair satisfying $\Phi$. We assume that also
the facts $R(t_{k+1}),\ldots,R(t_n)$ belong to $I$; otherwise we can
remove any negative literal from $\Phi$ if it is not in $I$. We also
recall that in the presence of one FD only, the family of 
globally-optimal and completion-optimal repairs coincide
(Proposition~\ref{prop:collapse}).
\begin{lemma}\label{lemma:global-test}
  A (completion-) globally-optimal repair $I'$ satisfying $\Phi$ exists if
  and only if the following conditions are satisfied:
\begin{description}
\item[(i)] $\{R(t_1),\ldots,R(t_k)\}$  is conflict-free;
\item[(ii)]
  $\{D_{R(t_1)},\ldots,D_{R(t_k)}\}\cap 
  \{D_{R(t_{k+1})},\ldots,D_{R(t_m)}\} = \varnothing$; 
\item[(iii)] $D_{R(t_j)}\cap\omega_{\mathord\succ}(C_{R(t_j)})\neq \varnothing$ for every 
  $j\in\{1,\ldots,k\}$.
\item[(iv)] $\omega_{\mathord\succ}(C_{R(t_j)})\setminus
(D_{R(t_{k+1})} \cup\ldots\cup D_{R(t_n)})\neq\varnothing$ for every 
$j\in\{k+1,\ldots,m\}$.
\end{description}
\end{lemma}
\begin{proof}
  For the \emph{only if} part, we take any globally-optimal repair
  $I'$ satisfying $\Phi$. \emph{(i)} and \emph{(ii)} are trivially
  satisfied. Assume that $I'$ is the result of
  Algorithm~\ref{alg:prioritized-repairing} with a choice sequence
  $R(s_1),\ldots,R(s_\ell)$. Take any $j \in \{1,\ldots,k\}$ and let
  $j'$ be the smallest index of a fact from $C_{R(t_j)}$ in the
  sequence. Clearly, $R(s_{j'})\in I'$. Since $R(t_j)$ also belongs to
  $I'$, both $R(s_{j'})$ and $R(t_j)$ belong to the same
  $(X,Y)$-cluster i.e., $R(s_{j'})\in D_{R(t_j)}$. Also prior to
  selecting $R(s_{j'})$ the temporary instance $I^o$ contains
  $C_{R(t_j)}$. Therefore
  $R(s_{j'})\in\omega_{\mathord\succ}(C_{R(t_j)})$ which proves {\em
    (iii)}.

  We show \emph{(iv)} similarly. For any $j\in\{k+1,\ldots,m\}$ let
  $j'$ be the smallest index of a fact from $C_{R(t_j)}$ in the
  sequence of choices used to construct $I'$. Prior to making the
  choice $R(s_{j'})$ the temporary instance $I^o$ contains
  $C_{R(t_j)}$, $R(s_{j'})\in\omega_{\mathord\succ}(C_{R(t_j)})$, and
  $R(s_{j'})$ does not belong to any of
  $D_{R(t_{k+1})},\ldots,D_{R(t_n)}$.

  For the \emph{if} part, we construct $I'$ using
  Algorithm~\ref{alg:prioritized-repairing} with a choice sequence
  $R(s_1),\ldots,R(s_\ell)$ defined as follows. By \emph{(i)} and {\em
    (iii)}, for $j\in\{1,\ldots,k\}$ the choice $R(s_j)$ is any fact
  from $D_{R(t_j)} \cap \omega_{\mathord\succ}(C_{R(t_j)})$. By {\em
    (ii)} and \emph{(iv)}, for any $j\in\{k+1,\ldots,m\}$ the choice
  $R(s_j)$ is any fact from
  $\omega_{\mathord\succ}(C_{R(t_j)})\setminus(D_{R(t_{k+1})}\cup\ldots\cup
  D_{R(t_n)})$. The remaining choices $R(t_j)$ for
  $j\in\{m+1,\ldots,\ell\}$ are selected in an arbitrary way. We
  observe that the first $k$ steps guarantees that the facts
  $R(t_1),\ldots,R(t_k)$ belong to the repair instance $I'$ (possibly
  placed there in later consecutive steps) and that $I'$ does not
  contain any of the facts $R(t_{k+1}),\ldots,R(t_m)$. \qed
\end{proof}
\begin{lemma}\label{lemma:pareto-test}
  A Pareto-optimal repair $I'$ satisfying $\Phi$ exists if and only if
  the following conditions are satisfied:
\begin{description}
\item[(i)] $\{R(t_1),\ldots,R(t_k)\}$ is conflict-free;
\item [(ii)]
  $\{D_{R(t_1)},\ldots,D_{R(t_k)}\}\cap\{D_{R(t_{k+1})},\ldots,D_{R(t_m)}\}=\varnothing$;
\item[(iii)] for every $j\in\{1,\ldots,k\}$, for every fact $R(t)\in
  C_{R(t_j)}\setminus D_{R(t_j)}$ there exists $R(t')\in D_{R(t_j)}$
  such that $R(t)\not\succ R(t')$.
\item[(iv)] for every $j\in\{k+1,\ldots,m\}$ there exists an
  $(X,Y)$-cluster $D$ of $C_{R(t_j)}$ different from
  $D_{R(t_{k+1})},\ldots,D_{R(t_m)}$ such that for every $t\in
  D_{R(t_{k+1})}\cup\ldots\cup D_{R(t_m)}$, there exists $R(t')\in D$
  such that $R(t)\not \succ R(t')$.
\end{description}
\end{lemma}
\begin{proof}
  For the \emph{only if} part, \emph{(i)} and \emph{(ii)} are trivially
  implied by $I'\models\Phi$. To show \emph{(iii)} and \emph{(iv)} we
  observe that a Pareto-optimal repair contains exactly one
  Pareto-optimal $(X,Y)$-cluster for every $X$-cluster. For clusters
  $C_{R(t_1)},\ldots,C_{R(t_k)}$ this together with the fact that
  $\{R(t_1),\ldots,R(t_k)\}\subseteq I'$ implies \emph{(iii)}. For
  clusters $C_{R(t_{k+1})},\ldots,C_{R(t_{m})}$ this together with the
  fact that $\{R(t_{k+1}),\ldots,R(t_m)\}\cap I' = \varnothing$
  implies \emph{(iv)}.

  For the \emph{if} part, we construct the repair $I'$ by selecting an
  $(X,Y)$-cluster from every $X$-cluster. Because Pareto optimality is
  defined in terms of neighboring facts and for one FD conflicts can
  be present only inside an $X$-cluster, to show that the repair $I'$
  is Pareto optimal it is enough to show that for every $X$-cluster
  the selected $(X,Y)$-cluster is Pareto optimal (among all
  $(X,Y)$-clusters in the $X$-cluster).

  For $X$-clusters $C_{R(t_1)},\ldots,C_{R(t_k)}$ we select
  $D_{R(t_1)},\ldots,D_{R(t_k)}$ resp. We note that by \emph{(i)} the
  $(X,Y)$-clusters belong to different $X$-clusters and by \emph{(ii)}
  we do not include any of the facts
  $R(t_{k+1}),\ldots,R(t_m)$. Pareto optimality is implied by {\em
    (iii)}. For $X$-clusters $C_{R(t_{k+1})},\ldots,C_{R(t_m)}$ we
  select the $(X,Y)$-clusters as described in \emph{(iv)}. Pareto
  optimality of those clusters is also implied by \emph{(iv)}. For an
  $X$-cluster other than $C_1,\ldots,C_m$ we select any
  $(X,Y)$-cluster that is Pareto optimal (for the $X$-cluster).  Since
  all selected $(X,Y)$-clusters are Pareto optimal, the instance $I'$
  is a Pareto-optimal repair such that $I'\models\Phi$.  \qed
\end{proof}

\begin{theorem}\label{thm:tractable-case}
  If the set of integrity constraints contains at most one functional
  dependency per relation name and no other constraints, then
  computing preferred consistent answers to quantifier-free queries is
  in PTIME for $\PRep$, $\GRep$, and
  $\CRep$.
\end{theorem}
\begin{proof}
  We adopt the algorithm from \cite{ChMa04}. We assume that the query
  $\Psi$ is in CNF i.e., $\Psi=\Psi_1\land\ldots{}\land\Psi_n$. By
  definition $\true$ is not a preferred consistent query answer to
  $\Psi$ if and only if there exists a preferred repair $I'$ and
  $i\in\{1,\ldots,n\}$ such that $I'\varnot\models\Psi_i$ i.e.,
  $I'\models\neg\Psi_i$. Note that the negation of $\Psi_i$ is a
  conjunction of literals. Consequently, the algorithm attempts to
  verify for every $i\in\{1,\ldots,n\}$ whether a preferred repair
  satisfying $\neg\Psi_i$ exists using tests from
  Lemma~\ref{lemma:global-test} or~\ref{lemma:pareto-test} (depending
  on the class of preferred repairs considered) If this condition is
  satisfied for some $i\in\{1,\ldots,n\}$, then $\true$ is not the
  preferred consistent answer to $\Psi$. On the other hand, $\true$ is
  the preferred consistent answer if the test fails for every
  $i\in\{1,\ldots,n\}$. Finally, we remark that the test can be
  performed in time polynomial in the size of the instance $I$ (the
  size of the query is assumed to be a constant) \qed
\end{proof}

\section{Related work}
\label{sec:related}
We limit our discussion to the work on using priorities to maintain
consistency and facilitate resolution of conflicts.

The first article to notice the importance of priorities in
information systems is \cite{FaUlVa83}. There, the problem of
conflicting updates in (propositional) databases is solved in a manner
similar to $\CRep$. The considered priorities are transitive, which is
more restrictive than acyclicity and does not bring any computational
benefits in our framework: our reductions can be modified to use only
transitive priorities. \cite{Br89} is another example of $\CRep$-like
prioritized conflict resolution of first-order theories. The basic
framework is defined for priorities which are weak orders. A partial
order is handled by considering every extension to weak order. This
approach also assumes transitivity of the priority.

In the context of logic programs, priorities among rules can be used
to handle inconsistent logic programs (where rules imply contradictory
facts). More important rules are satisfied, possibly at the cost of
violating less important ones. In a manner analogous to
Proposition~\ref{prop:global-alternative-definition}, \cite{NiVe02}
lifts a total order on rules to a preference on (extended) answers
sets. When computing answers only maximally preferred answers sets are
considered.


In \cite{Gr97}, Grosof presents a simpler approach to handling of
inconsistent logic programs with user priorities. Conflicting facts
are removed from the model unless the priority specifies how to
resolve the conflict. Because only programs without disjunction are
considered, this approach always returns exactly one model of the
input program. Constructing preferred repairs in a corresponding
fashion (by removing all conflicts unless the priority indicates a
resolution) would similarly return exactly one database instance
(fulfillment of $\mathcal{P}1$ and $\mathcal{P}4$). However, if the
priority is not total, the returned instance is not a repair and
therefore $\mathcal{P}5$ is not satisfied. Such an approach leads to a
loss of (disjunctive) information and does not satisfy $\mathcal{P}2$
and $\mathcal{P}3$.

In \cite{CaGrZu09}, Caroprese et al.\ propose the framework of
\emph{conditioned active integrity constraints}, which allows the user
to specify the way some of the conflicts created with a constraint can
be resolved. This framework satisfies properties $\mathcal{P}1$ and
$\mathcal{P}2$ but not $\mathcal{P}3$ and $\mathcal{P}4$. The authors
also describe how to translate conditioned active integrity
constraints into a prioritized logic program \cite{SaIn00}, whose
preferred models correspond to maximally preferred repairs.

In \cite{MoAnAc04}, Motro et al.\ use ranking functions on facts to
resolve conflicts by taking only the fact with highest rank and
removing others. This approach constructs a unique repair under the
assumption that no two different facts are of equal rank (satisfaction
of $\mathcal{P}4$). If this assumption is not satisfied and the facts
contain numeric values, a new value, called the fusion, can be
calculated from the conflicting facts (then, however, the constructed
instance is not necessarily a repair in the sense of
Definition~\ref{def:repair} which means a possible loss of
information).

In \cite{GrSiTrZu04}, Greco et al.\ study a different approach based on
ranking is studied. The authors consider polynomial functions that
are used to rank repairs. When computing preferred consistent query
answers, only repairs with the highest rank are considered. The
properties $\mathcal{P}2$ and $\mathcal{P}5$ are trivially satisfied,
but because this form of preference information does not have natural
notions of extensions and maximality, it is hard to discuss postulates
$\mathcal{P}3$ and $\mathcal{P}4$. Also, the preference among repairs
in this method is not based on the way in which the conflicts are
resolved.

In \cite{GrLe04}, Greco and Lenbo study an approach where the user has
a certain degree of control over the way the conflicts are
resolved. Using repair constraints the user can restrict considered
repairs to those where facts from one relation have been removed only
if similar facts have been removed from some other relation. This
approach satisfies $\mathcal{P}3$ but not $\mathcal{P}1$. A method of
weakening the repair constraints is proposed to get $\mathcal{P}1$,
however this comes at the price of losing $\mathcal{P}3$.

In \cite{AnFuMi06}, Andritsos et al.\ extend the framework of
consistent query answers with techniques of probabilistic
databases. Essentially, only one key dependency per relation is
considered and user preference is expressed by assigning a probability
value to each of mutually conflicting facts. The probability values
must sum to $1$ over every clique in the conflict graphs. This
framework generalizes the standard framework of consistent query
answers: the repairs correspond to possible worlds and have an
associated probability. We also note that no repairs are removed from
consideration (unless the probability of the world is $0$). The query
is evaluated over all repairs and the probability assigned to an
answer is the sum of probabilities of worlds in which the answer is
present. Although the considered databases are repairs, the use of the
associated probability values makes it difficult to compare this
framework with ours.

In \cite{GaSu10}, Gatterbauer and Suciu study the problem of conflict
resolution in the setting of community databases, where a group of
users, each having their own database over the same schema,
consolidate their knowledge of facts using mappings. This is
essentially a simplified peer-to-peer data exchange
setting~\cite{FKMT06}: the schema consists of one relation with a key
dependency and mappings permit to import facts not present in the
database of one user from the databases of other users. Facts imported
from different users may be conflicting and the authors propose to use
a total trust ordering on the mappings to resolve the conflicts. If
the mapping network is acyclic, then there exists a unique solution
(for every user), and in fact, this solution can be obtained using for
instance Algorithm~\ref{alg:prioritized-repairing} with a specially
precomputed, total priority and an instance containing the union of
all (accessible) facts. The main challenge addressed by the paper is
the setting where the mapping network is cyclic, which may yield
several solutions. Furthermore the users are allowed to specify
negative facts i.e., facts that are not believed to be true. These two
features render the setting incomparable with our approach: cyclic
mappings may possibly lead to a cyclic priority relation and conflicts
between negative and positive facts cannot be captured with denial
constraints.

\section{Conclusions and future work}
\label{sec:future}
In this paper we have proposed a general framework of preferred
repairs and preferred consistent query answers. We have also proposed a
set of desirable properties of a family of preferred repairs. We have 
presented three families of preferred repairs: $\PRep$,
$\GRep$, and $\CRep$ based on different notions of
optimality of compliance with the priority. For every repair family we
have presented a sound and complete database repairing
algorithm. Figure~\ref{fig:summary} summarizes the computational
complexity results; its first row is taken from \cite{ChMa04}.
\begin{figure*}
\begin{center}
\begin{tabular}{|c|c|c|c|}
\hline
   & \multirow{2}{*}{Repair Check} &
   \multicolumn{2}{c|}{Consistent Answers to} \\  
   \cline{3-4}
         &   & $\{\forall,\exists\}$-free queries 
         & conjunctive queries \\
\hline
\hline
   $\Rep$ & PTIME & PTIME & co-NP-complete \\
\hline 
$\GRep$ & ~co-NP-complete~ &
   \multicolumn{2}{c|}{$\Pi^p_2$-complete} \\
\hline
~$\PRep$~ & LOGSPACE & \multicolumn{2}{c|}{co-NP-complete} \\
\hline
$\CRep$ & PTIME & \multicolumn{2}{c|}{co-NP-complete} \\ 
\hline
\end{tabular}
\end{center}
\caption{\label{fig:summary} Summary of complexity results.}
\end{figure*}

We envision several directions for further work. We plan to
investigate other interesting ways of selecting preferred repairs with
priorities.  Also, extending our approach to cyclic priorities is an
intriguing and challenging issue. Including priorities in similar
frameworks of preferences \cite{GrLe04} leads to losing monotonicity
i.e., the property $\mathcal{P}5$ of the resulting family of preferred
repairs. A modified, conditional, version of monotonicity may be
necessary to capture non-trivial families of repairs. 

Along the lines of \cite{ABCHRS03}, the computational complexity
results could be further studied, by assuming the conformance of
functional dependencies with BCNF.  Finally, the class of constraints
can be extended to universal constraints~\cite{StCh08}. This class of
constraints allows to express conflicts caused not only by the
presence of some facts but also by simultaneous absence of other
facts. Conflict hypergraphs can be generalized to extended conflict
hypergraphs which include negative facts.


\end{document}